\documentclass[11pt,a4paper]{article}

\usepackage{jheppub}

\usepackage{multirow, graphicx,amssymb,url,mathrsfs,amsmath}
\usepackage{wrapfig,boxedminipage,setspace,subfigure,epsfig}
\usepackage{amsxtra,amstext,latexsym,dsfont,amsfonts}
\usepackage{color,eucal}
\usepackage[dvipsnames]{xcolor}
\usepackage{float}
\usepackage{slashed,comment}
\usepackage{kotex}
\usepackage{tensor}

\usepackage{indentfirst}

\newcommand{\ie}{{\it i.e.}}

\title{Entanglement structures from modified IR geometry}

\author[a]{Xin-Xiang Ju,}
\author[a]{Teng-Zhou Lai,}
\author[a]{Bo-Hao Liu,}
\author[b]{Wen-Bin Pan,}
\author[a,c]{and Ya-Wen Sun}

\emailAdd{juxinxiang21@mails.ucas.ac.cn}
\emailAdd{laitengzhou20@mails.ucas.ac.cn}
\emailAdd{liubohao16@mails.ucas.ac.cn}
\emailAdd{panwb@ihep.ac.cn}
\emailAdd{yawen.sun@ucas.ac.cn}

\affiliation[a]{School of Physical Sciences, University of Chinese Academy of Sciences, Zhongguancun east road 80, Beijing 100190, China}
\affiliation[b]{Institute of High Energy Physics, Chinese Academy of Sciences,\\19B Yuquan Road, Shijingshan District, Beijing 100049, China}
\affiliation[c]{Kavli Institute for Theoretical Sciences, University of Chinese Academy of Sciences, Beijing 100049, China}

\abstract{We investigate a new proposal connecting the geometry at various radial scales in asymptotic AdS spacetime with entanglement structure at corresponding real-space length scales of the boundary theory. With this proposal, the bulk IR geometry encodes the long-scale entanglement structure of the dual quantum system. We consider two distinct types of IR geometries, namely the spherical case and the hyperbolic case, which are intimately related to the physics of differential entropy and brane-world holography separately. We explore the corresponding change in the dual long-scale entanglement structures, utilizing the tools of the Ryu-Takayanagi formula, conditional mutual information, and partial entanglement entropy. The results indicate that modifying the IR geometry leads to a redistribution of entanglement at scales longer than a critical length determined by the location of the IR region, with the two modified IR geometries corresponding to two opposite ways of redistribution. Furthermore, we establish the maximum amount of entanglement that can be modified, which is proportional to the area of the IR region.
}

\begin{document}
\maketitle

\section{Introduction}
\noindent
The quantum informational properties of gravity have been playing a critical role in our understanding of the quantum aspects of gravity. It has long been recognized in the context of quantum gravity that quantum entanglement establishes a profound and intricate connection with geometry. This historical connection can be traced back to the seminal discovery of the Bekenstein-Hawking entropy of black holes  \cite{Bekenstein:1973ur,Bekenstein:1974ax,Bardeen:1973gs,Jacobson:2003wv,Eisert:2008ur} and the notion of the holographic principle \cite{tHooft:1993dmi,Susskind:1994vu,Maldacena:1997re}. The Ryu-Takayanagi (RT) formula \cite{Ryu:2006bv,Ryu:2006ef,Hubeny:2007xt,Nishioka:2009un} asserts that in the context of AdS/CFT, the entanglement entropy of a boundary spatial subregion is determined by a purely geometric quantity in the bulk: the area of the minimal surface homologous to the boundary subregion. This was further refined into the concept of subregion-subregion duality, which states that a bulk subregion enclosed by the RT surface could be fully reconstructed from operators in the corresponding boundary subregion \cite{Czech:2012bh,Wall:2012uf,Headrick:2014cta,Bousso:2022hlz,Espindola:2018ozt,Saraswat:2020zzf,dong2016reconstruction,Harlow:2018fse,Bousso:2012sj,Geng:2023iqd,Geng:2023qwm,Geng:2020fxl}. Developments in understanding the connection between quantum entanglement and geometry lead to the conception that ``entanglement builds geometry" \cite{Rangamani:2016dms,Swingle:2009bg,VanRaamsdonk:2009ar,VanRaamsdonk:2010pw}, inspiring the ER=EPR proposal \cite{Maldacena:2013xja}, which claims that quantum entanglement makes the geometry connected in spacetime. 

Another connection between quantum entanglement and geometry emerged through the development of the AdS/MERA (multi-scale entanglement renormalization ansatz) correspondence \cite{Swingle:2009bg,Vidal:2007hda}, which was later generalized to the continuous AdS/cMERA correspondence \cite{Nozaki:2012zj}. AdS/MERA correspondence states that an emergent asymptotic AdS spacetime could be defined from the amount of quantum entanglement of a quantum many-body system at renormalized real space scales. 

Building upon insights from MERA and some other facts which we will elaborate on later, in this paper, we take a step forward and propose a further connection between quantum entanglement and geometry. In contrast to the usual momentum space UV/IR relation in the dictionary, we propose that the geometry at different radial scales captures the structure of dual quantum entanglement at different real space scales on the boundary. To be more precise, the geometry near the boundary determines the short-length entanglement structure of the boundary degrees of freedom while the IR geometry encodes information of the long length entanglement structure of the boundary degrees of freedom. Here we emphasize that the short (long) length entanglement denotes the entanglement structure in the real space location of boundary degrees of freedom, in contrast to the UV (IR) entanglement, which we use to denote the momentum space physics\footnote{We distinguish between long (short)-range with long (short)-length here as in condensed matter physics, long-range entanglement specially refers to entanglement between degrees of freedom with infinite long distance in between, while here in the current context, we generally refer to entanglement at relatively long but still finite range. We also emphasize here that the long (short) distance refers to distance in the boundary while not the bulk distance.}. Therefore, modifying the IR geometry would result in a change of the long distance scale entanglement structure of dual boundary degrees of freedom. In other words, different IR geometries give different behaviors in the long-scale entanglement structures.

Note that this is different from the construction of AdS/MERA correspondence and the surface/state correspondence \cite{Miyaji:2015yva}, which was developed based on AdS/MERA, in the following sense. The construction of MERA emphasizes the role of renormalization in real space. 
In our proposal, we do not involve the renormalization procedure, and as a consequence, we do not eliminate the short-range entanglement. Additionally, the geometry in the bulk serves as the dual spacetime of the boundary quantum system in our case. This contrasts with MERA, where the AdS geometry is an emergent spacetime and not the original dual spacetime of the quantum system.
We claim that the entanglement structure at various scales, without being renormalized, could be encoded in the geometry at corresponding scales in the bulk. This offers deeper insights into the structure of entanglement in the real space beyond focusing only on the entanglement entropy. The connection between the bulk geometry at various radial scales and the corresponding real-space entanglement structures can be analyzed through the RT formula and fine entanglement structure measures.

Besides MERA and other real-space entanglement analysis \cite{Liu:2012eea}, the observations that inspired our proposal are the following. 
First, from the subregion-subregion duality, when we start from a small boundary subregion and enlarge it, more long-range entanglement structure of the boundary subsystem should be included in the larger boundary subregion and the corresponding bulk subregion would in general go deeper into the bulk. This implies that the inner region in the bulk should encode information about these additional longer range quantum entanglement of the dual boundary system. 

Second, the hole-ography proposed in \cite{Balasubramanian:2013rqa,Balasubramanian:2013lsa,Balasubramanian:2018uus} studied the entanglement entropy between a hole in AdS spacetime and its complement. The boundary state dual to this bulk spacetime with hole removed is a time band of the boundary. The boundary interpretation of this entanglement entropy is the so-called differential entropy \cite{Hubeny:2014qwa,Headrick:2014eia,Myers:2014jia,Czech:2014wka}. Differential entropy could be viewed as a state merging protocol of the boundary with short range entanglement structures kept unchanged while some of the long range entanglement structures removed due to the modified causal structure of the time band \cite{Czech:2014tva}. It is further proposed that the entanglement entropy could be a fine grained entropy associated with a dual boundary state whose certain long range entanglement structures have been removed, as could be evident from the concordance of observer physics on the two sides \cite{Ju:2023bjl,Ju:2023dzo}.  All this indicates that removing a hole in the IR of the bulk spacetime would correspond to removing certain long range entanglement structures in the boundary system. 

The subregion-subregion duality was also generalized to the subalgebra-subregion duality \cite{Leutheusser:2022bgi}, which states that the corresponding bulk spacetime subregion should be totally determined from the dual subalgebra associated with a given boundary subsystem. In the subalgebra-subregion duality, more general bulk spacetime subregions are studied beyond the usual entanglement wedges, especially including convex bulk spacetime subregions. The convexity condition could arise from the well defined condition of observers, where the complements of bulk convex subregions are proposed to be dual to boundary states with less long range entanglement \cite{Ju:2023bjl}. A simple example of a convex bulk spacetime subregion is also the spherical hole first studied in hole-ography, whose complement corresponds to a time band. The well-defined subalgebra and subalegbra-subregion correspondence for these convex subregions further support the consistency of the interpretation above.

In this paper, we investigate the consequence of this proposal, \ie, changing the IR geometry of a given spacetime and analyze the long-scale entanglement structures corresponding to different IR geometries. We modify the IR
geometry into two distinct types of geometries, namely the spherical case and the hyperbolic case. Utilizing the tools of RT formula, the conditional mutual information and the partial entanglement entropy \cite{Vidal:2014aal,Wen:2018whg,Wen:2019iyq,Wen:2020ech}, we could further show that different IR geometries indeed result in different long-scale entanglement structures. In this work, we will especially focus on extremal limits of the two types of modified IR geometries and more generalized cases will be investigated in a later work. We obtain the intriguing entanglement structures for those two extremal cases, which, inspiringly, exhibit opposite properties as their dual geometries do. The results in this work further support the proposal that IR geometry encodes information of the dual long scale entanglement properties.

Furthermore, in the sense of the equivalence of the entanglement structure, the spherical extremal case is closely related to ``hole-ography" \cite{Balasubramanian:2013lsa}, while the hyperbolic extremal case is connected to the existence of an end-of-the-world (EoW) brane in the bulk. Therefore, this work could further provide new perspectives on the quantum entanglement properties of these two systems through the lens of modifying long-range entanglement and the possibility to study the quantum information aspects of differential entropy, brane-world holography, and AdS/BCFT \cite{Fujita:2011fp,Takayanagi:2011zk} within a unified framework.

The remaining sections of this paper are organized as follows. In section 2, we introduce the basic setup of modifying the IR geometry and focus on two types of this modification. We then explore the details of the geometry and RT surfaces in both cases. In section 3, we analyze the fine entanglement structures of the modified geometries. Section 4 is devoted to conclusions and discussions.

\section{Modifying IR geometry}
\noindent
{To investigate the impact of modified IR geometries in the long-range entanglement structures of the dual quantum system, in this section, we introduce the basics of modifying IR geometries. Note that before modifying the IR geometry, the background would be the AdS spacetime, and the dual system would have a scale invariance, so that the entanglement structure between degrees of freedom at different distances would be the same. Modifying the IR geometry introduces an IR scale corresponding to the size of the IR region, thereby breaking the scale invariance of the entanglement structure. There are various possibilities to change the IR geometry and in this work, we will focus on two extreme cases where the physics becomes apparent, namely the spherical extremal case and the hyperbolic extremal case, respectively. We will first introduce the basic principles for modifying the IR geometries. Then we will analyze the geometrical structures and calculate the RT formula in these two extreme cases, providing the basis for the analysis of entanglement structures in the next section. We focus on the 2+1 dimensional bulk spacetime, while most of our results in this section are still valid in higher dimensions.}

\subsection{Basic setups and two general cases}
\noindent
In this paper, our objective is to compare the entanglement structures of boundary states corresponding to different bulk IR geometries. Various IR geometries in asymptotic AdS spacetime have been extensively explored within the context of AdS/CMT \cite{Zaanen:2015oix,Hartnoll:2018xxg}, including Lifshitz geometries \cite{Kachru:2008yh} and hyperscaling violating geometries \cite{Charmousis:2010zz}, to study different IR phases in strongly coupled quantum many-body systems. In that framework, the geometry flows from a specific IR geometry to the boundary asymptotic AdS geometry under irrelevant deformations in the IR. In this work, to distinctly isolate the long-scale entanglement structure and facilitate comparisons between entanglement structures of different IR geometries, unaffected by medium-scale geometries under the RG flow, we will not employ these types of geometries. Instead, we would like to glue an IR region with different IR geometries to a fixed surrounding AdS geometry, to ensure that the effects of the IR geometries are most visible.

To glue two distinct geometries, we adopt the brane-world scenario, i.e., the IR geometry and the UV part of the geometries are connected by an interface brane, with matter fields on the brane satisfying the junction conditions. This approach has been extensively investigated in the contexts of AdS/BCFT, AdS/ICFT, and brane-world holography.

In this section, we will first examine the general constraints for possible IR geometries, subsequently selecting two classes of IR geometries that are most interesting for the long-range entanglement structure analysis. In the most generalized case, the spacetime that we consider does not need to be static so as to accommodate as many interesting cases as possible. We demonstrate the guaranteed existence of these geometries and investigate their entanglement properties. The details of these explicit solutions will be left to future work.

{To achieve this, we will first pick a Cauchy slice in the bulk spacetime and choose a bulk IR subregion. We will manipulate the geometry inside this IR region on the Cauchy slice (IR geometry) while keeping the geometry outside of the region (UV geometry) unchanged.} {This could be done by adding matter fields that obey energy conditions at the edge of the IR region, which could be viewed as a matter brane. We need connection conditions for the IR and the outside geometries. Note that as we are working on a Cauchy slice, we only connect the spatial part of the two manifolds and let them evolve in time following Einstein equations. {Therefore, we do not utilize the Israel junction condition, which is for the Lorentz manifold while we have a 2d Riemann manifold here on a Cauchy slice. The condition that we need is to have the same induced metric (length as the edge is 1d) from the two sides at the connection edge {and the matter fields on the brane need to satisfy the Einstein equations on the brane.}
}

We demand the Cauchy slice always have zero extrinsic curvature, \ie{}, the modification is ``intrinsic" rather than a diffeomorphism of rechoosing a bulk slice with the same boundary. In this way, the HRT formula could be applied simply as will be shown later. We have various choices for the IR geometries, and we can even change the topology of the IR region, such as creating a wormhole that connects it to the Cauchy slice of another asymptotic AdS space. On the dual field theory side, this corresponds to entangling two CFTs together \cite{Maldacena:2013xja}.

Before analyzing the energy-momentum of the matter to be placed, we could have some observations about general properties of possible IR geometries from the Gauss-Bonnet theorem. The Gauss-Bonnet theorem signifies a common property of any consistent 2d Riemann manifold $M$, which we take to be the IR region here. The Gauss-Bonnet theorem is then given by
\begin{equation}\label{GB}
    \frac{1}{2}\int_{M} {dx^2} \sqrt h R_c + \oint_{\partial {M}} k_g ds =2 \pi \chi(M),
\end{equation}
where {the first term is the integration of ${R_c}$, the {intrinsic} curvature scalar} of the Cauchy slice $M$, $k_g$ is the geodesic curvature of {the boundary of the IR region} of the IR region, and $\chi(M)$ is the Euler characteristic number of the IR region. 
If we fix the edge and the topology of the IR region, equation \ref{GB} informs us that changing the interior IR geometry while keeping the outside geometry fixed can be viewed as a redistribution of the Cauchy slice curvature ${R_c}$ inside that IR region.

{Possible modified IR geometries are constrained by the Gauss-Bonnet theorem above, and as it is expected that the Ricci scalar $R_c$ reflects the geometry property which determines the behavior of dual quantum entanglement, we could consider the modification of IR geometries from the distribution of $R_c$.} We focus on the radial redistribution of $R_c$ and there are then two primary directions for this redistribution within the IR region
\begin{itemize} 
    \item{Increasing the curvature scalar in the center of the IR region while decreasing it near the edge of the IR region;}
    \item{Decreasing the curvature scalar in the center of the IR region while increasing it near the edge of the IR region.}
\end{itemize}
Here, ``near" refers to a thin shell whose thickness can be ignored, located near the edge of the IR region. {Those two possibilities lead to two extremal geometries as follows}
\begin{itemize} 
    \item{I. The spherical extremal case: the curvature scalar inside the IR region becomes very large so that any geodesic will not enter this region. The whole IR region becomes an entanglement shadow. The modified IR region is called spherical because it has positive curvature here.}
    \item{II. The hyperbolic extremal case: the curvature scalar inside the IR region tends to negative infinity so that any geodesic length inside the IR region tends to zero. The name hyperbolic refers to the negative curvature IR region.}
\end{itemize}
{We will explain these statements in more detail later.}
The motivation for analyzing these two extremal cases is as follows: it is expected that these extremal IR geometries would correspond to extremal long-range entanglement structures of the dual system. Therefore, the physics would be more evident in these cases. Other geometries with the same IR topology, including the unmodified original IR region, are dual to intermediate states between them, which we will discuss in a forthcoming paper \cite{Ju:2024}. In the following sections, we will focus on the two extremal cases above.

After designing the IR geometry of an AdS Cauchy slice, we set the initial extrinsic curvature of this Cauchy slice to zero, and evolve it in both directions in time to construct a complete bulk spacetime. This could always be done, as a standard process in numerical general relativity. {During this process, the non-zero energy-momentum tensor of matter fields within the IR region satisfies} the Einstein constraint equations
\begin{equation}\label{EinCon}
    \begin{aligned}
        R_c-K^{\mu\nu}K_{\mu\nu}+K^2 &=16\pi G \mu, \\
        \mathcal{D}^\mu K_{\mu \nu}-\mathcal{D}_\nu K  &=8\pi Gj_\nu , 
    \end{aligned}
\end{equation} at each Cauchy slice, {where the matter energy and momentum densities are determined as $\mu=\frac{R_c}{16\pi G}$ and $j_\nu=0$ separately when the extrinsic curvature $K_{\mu\nu}$ of the Cauchy slice is set to $0$.} We demand that the null energy condition (NEC) is valid while the weak energy condition (WEC) could be violated. The former stipulates $\mu + p_i \geq 0$, while the latter further stipulates the positivity of $\mu$. Specific examples of regular geometries can be constructed that are extremely close to\footnote{The two extremal geometries are singular at the exact limiting point, therefore we could employ geometries which are extremely close to the singular geometries but still are regular.} the two extremal cases mentioned above, which satisfy NEC while violating WEC. {This is because when redistributing $R_c$, we must use some matter with positive $\mu$ and some with negative $\mu$.} For static bulk spacetimes, we can directly use the RT formula on this Cauchy slice. Furthermore, in principle, we permit the geometry to be dynamical, and in that case, we would have to use the HRT formula \cite{Hubeny:2007xt} instead. The minimal surfaces (geodesics) on the $K_{\mu\nu}=0$ Cauchy slice would give HRT surfaces with two zero null extrinsic curvatures \cite{Rangamani:2016dms}. With the null energy condition satisfied, {the uniqueness of HRT surface} can be proved \cite{Ju:2023bjl}.

\subsection{The spherical modified case}
\noindent
In this subsection, we will analyze the geometrical structure of the first extremal case listed above, which we name `the spherical extremal case.' As we will demonstrate below, in this case, the IR region becomes an entanglement shadow \cite{Balasubramanian:2014sra}, where no RT surfaces can penetrate it.

\subsubsection{Geometry and RT surface}
\noindent
As we stated, the first way to modify the IR geometry is to redistribute the curvature on the Cauchy slice by increasing the curvature scalar in the center of the IR region while decreasing it near the edge of the IR region. For a spherically symmetric IR region, the procedure could be understood as using a spherical crown {(hemisphere in the extremal limit)} with a constant positive curvature scalar {to replace the IR geometry.} {The radius of the spherical crown has to be no less than the radius of the} {edge of the IR region} {in order to glue the two edges together consistently geometrically.} {The extremal case corresponds to the limit when the crown becomes a hemisphere. In this case,} the edge of the IR region is the equator of the hemisphere.

{For simplicity, we take} {a $K_{\mu\nu}=0$ Cauchy slice in a static spherically symmetric 2+1 dimensional spacetime} as an example. The ansatz for the metric for the whole spacetime is
\begin{equation}\label{ansatz}
ds^2=-f(r)g(r) dt^2+\frac{1}{g(r)}dr^2+r^2d\theta^2.
\end{equation}
We will pick the $t=0$ Cauchy slice, which has zero extrinsic curvature as the spacetime is static. The intrinsic curvature scalar of the Cauchy slice is given by
\begin{equation}\label{RC}
R_c=-\frac{g'(r)}{r}.
\end{equation}
We choose the spatial components of the metric to be the usual spatial parts of global AdS outside the IR region and a spherical metric inside the IR region
\begin{equation}\label{g(r) 1}
g(r)=\begin{cases}
(l^2-r^2)/l^2, & \text{for } r<r_{IR},\\
(l_{AdS}^2+r^2)/(l_{AdS}^2), & \text{for } r>r_{IR},
\end{cases}
\end{equation} 
where $r_{IR}$ is the radial location of the gluing edge and the radius of the spherical crown in the IR region is $l$. $l$ is required to be larger than $r_{IR}$ so that the UV region could be glued to the IR spherical crown at a latitude whose circumference of the boundary circle is the same as the edge. {The metric outside the IR region is still the pure $AdS_3$ geometry, while inside the IR region, in general, the metric needs to be supplemented by appropriate matter fields to satisfy the Einstein equations. In other words, the matter fields inside the IR region should be appropriately chosen to satisfy the Einstein equations as determined by the IR metric in (\ref{g(r) 1}).}

{{Simultaneously, matter fields also need to be present at the edge of the IR region, determined by the Einstein equations nearby the edge, \ie, by the junction conditions.} The discontinuity of $g(r)$ at $r=r_{IR}$ implies a divergence in $R_c$, meaning a matter ring with divergent $\mu$ located on the gluing edge, which is a common feature in brane-world scenario.} Note that the connection condition requires that the tangential component of the metric needs to be the same at both sides of the gluing edge, which is the same $g_{\theta\theta}=r^2$ term in this case. We emphasize here again that we are considering an IR geometry supported by matter fields at the edge of the IR region, which could be viewed as a brane. This scenario is analogous to the AdS/ICFT case, where the left and right bulks differ from each other yet can be connected through a specific profile at an interface brane which can be achieved by introducing an appropriate dynamic, brane-localized scalar field, e.g. in \cite{Liu:2024oxg}. The corresponding connection conditions in this case are also satisfied.

In the spherical extremal case, we have $l=r_{IR}$ {where the IR geometry becomes a hemisphere, and the entire IR region becomes an entanglement shadow as will be shown later.} Note that when $l$ is exactly $r_{IR}$ the metric would possibly contain a horizon at $r_{IR}$, therefore, we consider $l$ to be larger but very close to $r_{IR}$ case as the limiting case of the spherical extremal case so as to avoid possible complexities as mentioned above.

{The null energy condition gives
\begin{equation}
    \begin{aligned}
        &f'(r)\geq 0,\quad \text{and} \\
        &-r g(r) f'(r)^2+f(r) (3 r f'(r) g'(r)+2 r g(r) f''(r))-2 f(r)^2 (g'(r)-r g''(r))\geq 0,
    \end{aligned}
\end{equation}}
which could be satisfied in most cases by choosing a suitable function $f(r)$. {Let us examine this condition more closely. If we choose $f(r)$ to be a trivial function of $1$, the second condition above would imply monotonicity of $R_c(r)$, but that is not the case here based on (\ref{RC}) and (\ref{g(r) 1}). Therefore, to satisfy the null energy condition, $f(r)$ needs to be finely tuned, which might not be easily achievable. Physically, this is because $\mu$ diverges negatively at the edge, unlike most other systems where $\mu$ diverges positively at the brane. Hence, meeting the null energy condition $\mu+ p\geq 0$ is challenging. One possible solution is to extend the edge into a thin shell so that the matter does not exhibit negative divergence, and it is easier to find an $f(r)$ as long as the shell is thick enough. Another option is to relax the null energy condition, which we initially required to ensure uniqueness of the HRT surface. However, since we are considering a static spacetime here, NEC doesn't need to be present to preserve this property.}

{Now we calculate the RT surface in this background with modified IR geometry of a general spherical crown replacement, including but not limited to the spherical extremal case. The results will apply to both the nonextremal and extremal cases. We will solve for the geodesic in the modified geometry and compare it with the original geodesic for the same boundary subregion in the original spacetime. When the boundary subregion AB is small enough, the geodesic would stay the same as it will not reach the IR region. Therefore, we consider the case when AB is large enough for the geodesic to touch the IR region and at the same time smaller than half space without loss of generality.} 

{Consider a geodesic $\Gamma_{AB}$ with two endpoints A and B on the asymptotic boundary. When the length of the boundary interval $L_{AB}$ is large enough, $\Gamma_{AB}$ will intersect the edge of the IR region at points C and D. In this case, $\Gamma_{AB}$ will be divided into three parts $\Gamma_{AC}$, $\Gamma_{CD}$ and $\Gamma_{DB}$, and each of them is a geodesic in their respective region. From (\ref{g(r) 1}) we can see that $\Gamma_{CD}$ is a geodesic on a 2-sphere with radius $l$, which is part of the great circle of the sphere. Introducing a new parameter $\phi=\arcsin{\frac{r}{l}}$, the equation for $\Gamma_{CD}$ is}
\begin{equation}
    \cot{\phi}=a_1\cos{(\theta-\theta_1)},
\end{equation}
where $a_{1}$ and $\theta_1$ are integral constants that could be determined from the location of $C$ and $D$. We choose C and D to be symmetric with respect to the polar axis $\theta=0$ for simplicity, which gives $\theta_C=-\theta_D$ and $\theta_1=0$. 

$\Gamma_{AC}$( and $\Gamma_{BD} $) is the geodesic on the outside 2-hyperboloid. Introducing a new parameter $\psi$, $\psi=\sinh^{-1}\frac{r}{l_{AdS}}$, the equation for $\Gamma_{AC}$ is
\begin{equation}
    \coth{\psi}=a_2\cos{(\theta-\theta_2)},
    \label{great circle}
\end{equation}
where $a_2$ and $\theta_2$ are again integral constants. Similarly, $\Gamma_{BD} $ has  \begin{equation}
    \coth{\psi}=a_3\cos{(\theta-\theta_3)},
\end{equation} and from the same symmetry that sets $\theta_1=0$, here we have $a_2=a_3$ and $\theta_3=-\theta_2$.

The continuity and smoothness of the geodesic \cite{Anous:2022wqh} {(e.g. in section 3.3)} require that
\begin{equation}
   \lim_{r \to r_{\text{IR}}^+} \frac{dr}{d\theta}\bigg|_{C(D)}=\lim_{r \to r_{\text{IR}}^-}\frac{dr}{d\theta}\bigg|_{C(D)}.
    \label{c and s}
\end{equation}
The deepest point P, which has the global minimum of $r$, is given by 
\begin{equation}
    \cot{\phi_P}=a_1,
    \label{P modified}
\end{equation}
where $r_P=l r \sin{\phi_P}$. Using (\ref{great circle}) to (\ref{P modified}), we can find the boundary length of boundary interval AB 
 \begin{equation}
     L_{AB}=2l_{S^1}(\theta_2+\arccos{\frac{1}{a_2}}),
 \end{equation}
where $2\pi l_{S^1}$ is the spatial size of the boundary system, $\theta_2$ and $a_2$ are
\begin{equation}
a_2=\frac{\sqrt{2l^2l_{AdS}^2r_P^2+l^2 r_P^2 r_{IR}^2+l_{AdS}^4(l^2+r_P^2-r_{IR}^2)}}{l r_P\sqrt{l_{AdS}^2+r_{IR}^2}},
\end{equation}
\begin{equation}
    \theta_2=\arccos{\frac{r_P\sqrt{l^2-r_{IR}^2}}{r_{IR}\sqrt{l^2-r_P^2}}}-\arccos{\frac{\sqrt{l_{AdS}^2+r_{IR}^2}}{a_2}}.
\end{equation}

Now we compare this configuration of geodesic with the one with the IR geometry unchanged. For the same boundary subregion AB, we could compare the two geodesics with and without the IR geometry modified, especially their difference in shape and size. The conclusion is that in the IR modified geometry of this spherical crown case, the length of the geodesic for the same boundary subregion AB would be larger in the modified geometry and smaller in the original geometry. The geodesic does not reach as deep into the IR as in the original geometry and is ``repelled" towards the UV. The first point could easily be found from the fact that the length in the modified geometry becomes larger due to the new metric. 

To show the second point, technically we should start from the same boundary subregion AB and compare the two geodesics in the modified and unmodified geometries to see which one is closer to the UV. This is a technically quite involved task because in the modified geometry to find $r_P$ from AB is more difficult than the other way round, \ie{} to start from $r_P$ and find the corresponding AB. Thus here we take this way. {Consider a geodesic 
$\Gamma_{A'B'}$ in the original geometry with the deepest point P', which has the same $r$ value as P, $r_{P'}=r_P$.} Its boundary length is 
\begin{equation}
    L_{A'B'}=2l_{S^1}\arccos{\frac{1}{\coth{\psi_P}}},
    \label{P and boundary}
\end{equation}
where $\psi_P=\sinh^{-1} \frac{r_P}{l_{AdS}}$. One can check that $L_{A'B'}<L_{AB}$ when $r_{IR}<l<l_{AdS}$. Conversely, {given a boundary interval, the $r$ value of the deepest point in the unmodified case is always less than the modified case.} Note that this argument has a subtle point as to whether the same $r$ value of P is meaningful as the IR geometry has changed.

Nevertheless, this repelling behavior could be intuitively explained through pure geometric observation, confirming its validity. The left side of figure \ref{RTM} shows the shapes of geodesics in the spherical modified and unmodified geometry. $A, A', B'$ and $B$ are points on the boundary. Black dotted lines are geodesics in the original unmodified AdS geometry while the blue curve is the geodesic in the modified geometry. The pink region denotes the modified IR region. 
It is plotted using the stereographic projection, which can be viewed as a conformal map preserving the smoothness of the blue geodesic at points $C$ and $D$. In the stereographic projection of the spherical crown, the geodesic is an arc $\Gamma_{CD}$ being part of the blue dashed circle. For the same turning point P in the modified and unmodified geometry in this stereographic projection, the geodesic in the original spacetime is $\Gamma_{A'B'}$ while the geodesic in the modified geometry is $\Gamma_{AB}$. For the same points $AB$, the geodesic in the original spacetime would go deeper in the IR as shown in the figure. The intuitive explanation for this repelling behavior is as follows. When the IR geometry gets modified, the original geodesic, the black dashed $\Gamma_{AB}$ does not have the minimum length anymore. Inside the IR region, we could substitute the original IR geodesic black dashed $\Gamma_{EF}$ by an IR geodesic in the modified geometry, which is the blue dashed curve $\Gamma_{EF}$. Then this new curve $\Gamma_{AEFB}$ is shorter than the original one. However, this is still not the one with least length because the connection is not smooth at $E$ and $F$. Thus we could fine-tune near the point $E$ so that the curve becomes smoother, and the total length becomes shorter, utilizing the curved space version of the fact that straight line segment is the shortest between two fixed points. This fine-tuning always takes the curve closer to the UV due to the shape of $\Gamma_{EF}$ (blue dashed).

\begin{figure}[H]
    \centering 
\includegraphics[width=0.9\textwidth]{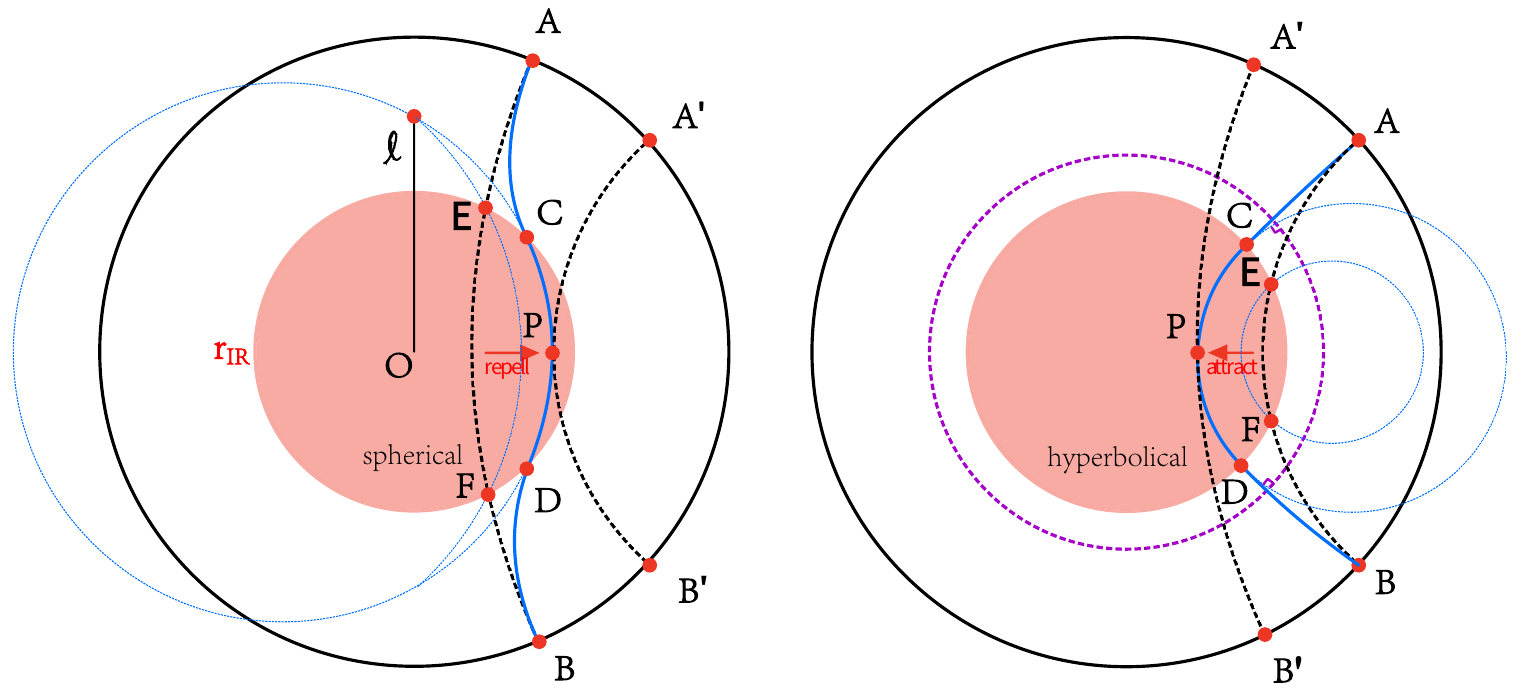} 
   \caption{Shapes of geodesics in the spherical (left) modified geometry and the hyperbolic (right) modified geometry. In the spherical modified case, RT surfaces tend to be ``repelled" away from the IR region (pink shaded region), while in the hyperbolic modified case, RT surfaces would be ``attracted into" the IR region. The figures are plotted using the stereographic projection. In both figures, the black dashed curves are RT surfaces in the original AdS vacuum geometry, and the blue curves are RT surfaces in the IR modified geometry. The blue dashed arcs inside the IR region represent the geodesics in the IR region, being part of the great blue circles. In the spherical extremal case $l=r_{IR}$, the blue circle coincides with the boundary of the IR region, and geodesic $\Gamma_{CD}$ will be glued onto it. In the hyperbolic extremal case, the purple circle (fictitious conformal boundary of the IR hyperbolic region) coincides with the boundary of the IR region, and geodesic $\Gamma_{CD}$ will be perpendicular to it at points $C$ and $D$.}\label{RTM}
\end{figure}

Therefore, the geodesic is ``repelled" {towards the UV region} in the spherical crown modified geometry. An illustration for this repelling behavior is shown in figure \ref{RTM}. In the extremal case when $l=r_{IR}$, the edge of the IR region becomes the equator, which is already the minimal geodesic. Therefore, the geodesic will never penetrate the IR region in this spherical extremal case. The quantum information theoretic explanation for this system, especially the physical implication of the repelling, will be presented in the next section. 

\subsubsection{Phase transition of RT surfaces in the spherical extremal case}

\noindent
{From here on, we focus on the extremal case in the spherical replacement case.} The IR region becomes an entanglement shadow \cite{Balasubramanian:2014sra} where no RT surfaces can penetrate into. As one enlarges the boundary subregion, the shapes of its RT surfaces may transition through four distinct ``phases" as follows. 

I. For a boundary subregion with a length $L$ small enough, its RT surface will fully reside in the outside of the IR region and does not touch the edge of the IR region. In this case, the RT surface in the modified geometry would stay the same as in the original AdS spacetime. {There exists a critical length $L_c$ and when the size of the boundary subregion $L$ reaches $L_c$, the RT surface will be tangent to the edge of the IR region. The deepest point P has $r_P=r_{IR}$ at the critical point. According to (\ref{P and boundary}), the critical length $L_c$ is}
\begin{equation}\label{deepest}
    L_c=2l_{S^1}\arccos{\frac{r_{IR}}{\sqrt{r_{IR}^2+l_{AdS}^2}}}=2l_{S^1}\arctan{\frac{l_{AdS}}{r_{IR}}}.
\end{equation}

II. For a boundary subregion with a length $L_c<L<\pi l_{S^1}$, the RT surface will wrap around the IR region without penetrating it, as shown in figure \ref{CMI}\footnote{Rigorously speaking, according to Hilbert's theorem, the UV geometry, being hyperbolic, cannot be globally embedded into the high-dimensional Euclidean geometry. Therefore, figure \ref{CMI} serves solely as an explicit illustration for embedding the IR geometry.}.
\begin{figure}[H]
    \centering 
   \includegraphics[width=0.8\textwidth]{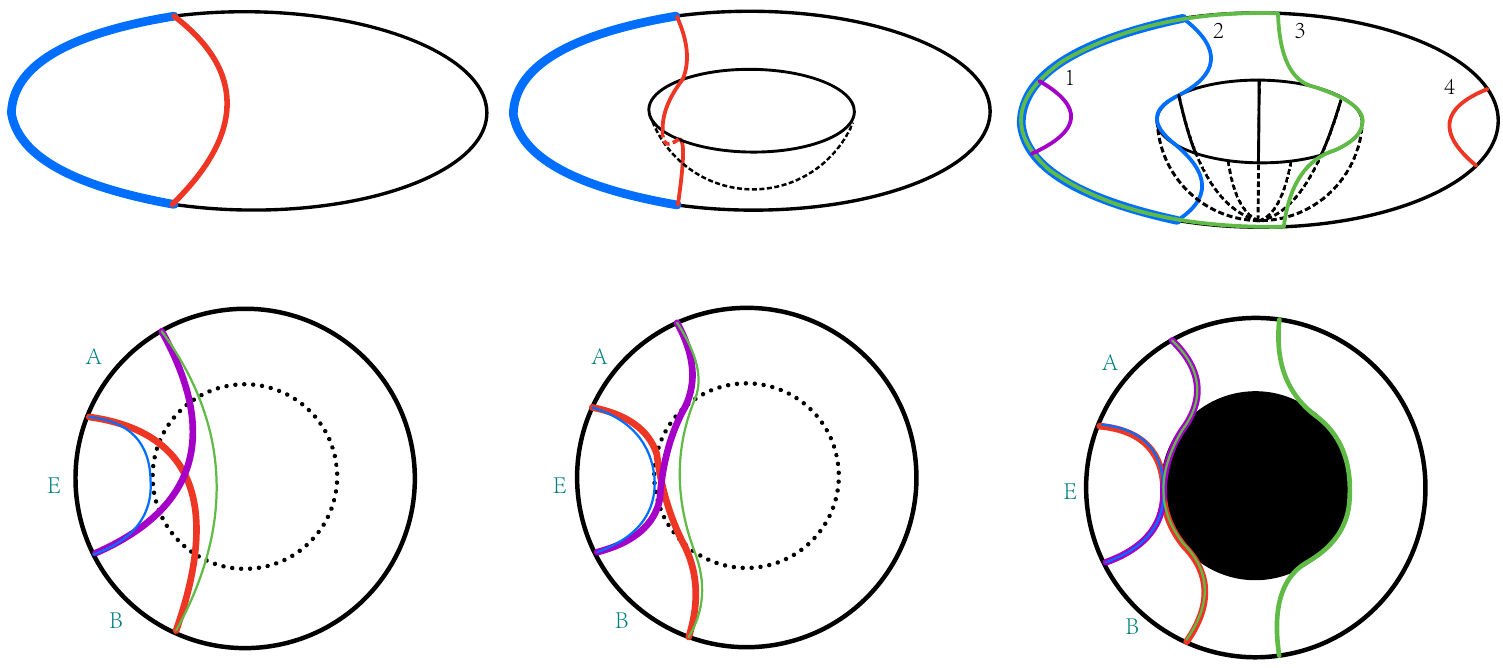} 
   \caption{Comparison of the RT surfaces in  the spherical modified IR geometry with the unmodified geometry. The figures on top illustrate Cauchy slices embedded in a high-dimensional Euclidean geometry, while the figures below show the RT surfaces on them plotted in stereographic projection. 
   In the left figures, we have the original pure AdS bulk geometry. In the middle figures, the original geometry inside the black dashed circle is replaced with a region of positive curvature, causing the geodesics to be ``repelled" from the center of AdS and bringing the boundaries of the entanglement wedges of subregions $AE$, $BE$, $E$ and $ABE$ closer to each other. The right figures represent the spherical extremal case, where RT surfaces are entirely repelled out of the IR region and the IR region becomes an entanglement shadow. The four different phases for RT surfaces for the spherical extremal case are illustrated in the top right figure.}
    \label{CMI} 
\end{figure}

III. For a boundary subregion with a length $\pi l_{S^1}<L<2\pi l_{S^1}-L_c$, its RT surface is the same as the RT surface of its complement, which is in the second phase. The entanglement wedge of this boundary subregion now includes the IR region. 

IV. For a boundary subregion with a length $L>2\pi l_{S^1}-L_c$, its RT surface does not touch the edge of the IR region, which is the same as in vacuum AdS spacetime.

The entanglement entropy of a boundary subregion in the spherical extremal case is then summarized as follows:
\begin{equation}\label{SPH}
    S(L)=
        \left\{
        \begin{aligned}
        &\frac c3 \log(\frac{l_{S^1}}{\pi\epsilon}\sin(\frac{L}{2l_{S^1}})),\,\,\,\,\,\quad\quad\quad\quad\quad\quad\quad\quad\quad\quad\quad\quad\quad(L<L_c\,\text{or}\,L>2\pi l_{S^1}-L_c)\\
        &\frac c3 \log(\frac{l_{S^1}}{\pi\epsilon}\sin(\frac{L_c}{2l_{S^1}}))+\frac {c\cot(L_c/2l_{S^1})}{6l_{S^1}}(L-L_c) ,\,\,\quad\quad\quad\quad\quad\quad\quad(L_c\leq L\leq \pi l_{S^1})\\
        &\frac {c}3 \log(\frac{l_{S^1}}{\pi\epsilon}\sin(\frac{L_c}{2l_{S^1}}))-\frac {c\cot(L_c/2l_{S^1})}{6l_{S^1}}(L-2\pi l_{S^1}+L_c), \,( \pi l_{S^1} \leq L\leq 2\pi l_{S^1}-L_c)
        \end{aligned}
        \right.
\end{equation}
where $l_{S^1}$ is the radius of the conformal boundary $S^1$.

\subsection{The hyperbolic modified case}
\noindent
{In the spherical extremal case, $\lim_{r\to r_{IR}} g_{rr}\to \infty$, preventing any RT surface from penetrating into the IR region. Conversely, in the hyperbolic extremal case, $g_{rr}\to 0$ when $r<r_{IR}$. Consequently, RT surfaces tend to penetrate the IR region perpendicularly whenever possible, leading to intriguing physical implications.} In this subsection, we will analyze the geometrical structure and the RT surfaces of the hyperbolic modified case, especially emphasizing the hyperbolic extremal case.

\subsubsection{Geometry and RT surfaces}
\noindent
Using the same ansatz (\ref{ansatz}), in the hyperbolic case, we have
\begin{equation}\label{Hy}
g(r)=\begin{cases}
(l^2+r^2)/l^2, & \text{for } r<r_{IR},\\
(l_{AdS}^2+r^2)/l_{AdS}^2, & \text{for } r>r_{IR},
\end{cases}
\end{equation}
{where $0<l<l_{AdS}$ is the ``radius" of the hyperbolic space inside the IR region. The curvature scalar inside the IR region $R\propto -1/l^2$ tends to the negative infinity when $l\to 0$, which gives the hyperbolic extremal case.}

Certain matter fields will be present in the IR region for the metric to be a solution of the Einstein equation. The matter fields and energy conditions will be similar to the first case, which we do not repeat here. Now we calculate a general RT surface in this modified background. We consider a large enough boundary region $AB$, and the corresponding RT surface again consists of three parts with two intersection points C and D on the edge of the IR region. $\Gamma_{CD}$ follows
\begin{equation}
    \coth{\phi}= a_1 \cos{\theta},
\end{equation}
where the new parameter $\phi=\sinh^{-1} \frac{r}{l}$ and $a_1$ is an integration constant. {The equation of $\Gamma_{AC}$ (or $\Gamma_{BD})$ is given by}
\begin{equation}
    \coth{\psi}=a_2\cos{(\theta-\theta_0)},
\end{equation}
where $\psi=\sinh^{-1} \frac{r}{l_{AdS}}$ and $a_2, \theta_0$  are integration constants. All integration constants can be determined from the coordinates of endpoints $A$ and $B$. Continuity and smoothness again require equation (\ref{c and s}) to hold at points C and D.

Consider a geodesic with the deepest point P, $r_P=l\sinh{\phi_P}=l_{AdS}\sinh{\psi_P}$. The length of the boundary region $AB$ is
\begin{equation}
    L_{AB}=2l_{S^1}(\theta_0+\arccos{\frac{1}{a_2}}),
\end{equation}
where $\theta_0$ and $a_2$ are
\begin{equation}
    a_2=\frac{\sqrt{2l_{AdS}^2l^2r_P^2+l^2r_P^2r_{IR}^2+l_{AdS}^4(l^2-r_P^2+r_{IR}^2)}}{rl\sqrt{l_{AdS}^2+r_{IR}^2}},
\end{equation}
\begin{equation}
    \theta_0=\arccos{\frac{r_P\sqrt{l^2+r_{IR}^2}}{r_{IR}\sqrt{l^2+r_P^2}}}-\arccos{\frac{\sqrt{l_{AdS}^2+r_{IR}^2}}{a_2}}.
\end{equation}
The length of the boundary region in the unmodified case is still \ref{P and boundary}. This time, $L_{A'B'}>L_{AB}$, which is the opposite of the spherical IR case. Therefore, when the boundary regions are the same, the deepest point in the hyperbolic case is deeper than the unmodified case. The RT surface will be ``attracted" to the IR region, as shown in the right side of figure \ref{RTM}.

\subsubsection{Phase transition of RT surfaces of the hyperbolic extremal case}
\noindent  In the hyperbolic extremal limit of $l\to 0$ in (\ref{Hy}), the IR region tends toward a ``light cone" in the embedded Minkowski spacetime (as depicted on the right side of figure \ref{IR-}), {with any geodesics inside the IR region having zero length}\footnote{In fact, it is crucial to maintain classical geometry, meaning the geodesic length within the IR region must satisfy $l_P \ll l \ll l_{AdS}$, where $l_P$ is the Planck scale. However, this minor adjustment to the area of the RT surface scales proportionally with $G$, but it does not affect the leading order term, which is our primary focus and scales as $O\left(\frac{1}{G}\right)$.}. This implies that the boundary of the IR region becomes equivalent to an EoW brane in the sense of the interpretation of the holographic entanglement entropy. In this extremal limit, the IR geometry is in fact the trivial spacetime proposed in \cite{Miyaji:2014mca}, in the sense that it has no quantum entanglement. For RT surfaces homologous to the boundary, penetrating into the IR region ``as fast as possible" is an optimal way to reduce their area. Consequently, if RT surfaces intersect with the boundary of the IR region, they must be perpendicular to it. As one enlarges the boundary subregion, the shapes of its RT surfaces may transition through three distinct ``phases" as follows.

I. For a boundary subregion with a length $L$ small enough, its RT surface will fully reside outside the IR region and does not touch the edge. In this case, the RT surface in the modified geometry remains the same as in the original AdS spacetime.

II. As the boundary subregion enlarges to $L=L_c$, its RT surface will undergo a first order phase transition before reaching the edge of the IR region (the geometric determination of $L_c$ will be provided below). After this transition, the RT surface will penetrate into the IR region and the part inside the UV region becomes two separate geodesics, each perpendicular to the boundary of the IR region, as illustrated in figure \ref{IR-}.

III. For a boundary subregion with a length $L > 2\pi l_{S^1} - L_c$, its RT surface does not touch the edge of the IR region, resembling the behavior in vacuum AdS spacetime. The entanglement wedge of this subregion contains the IR region. 

\begin{figure}[H]
    \centering \includegraphics[width=0.8\textwidth]{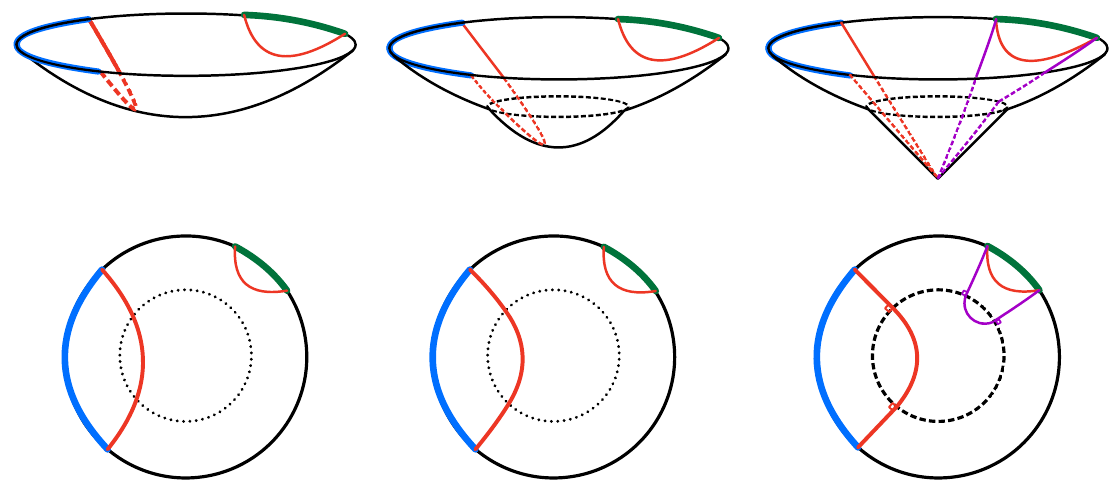} 
    \caption{Comparison of the RT surfaces in the hyperbolic modified IR geometry with the unmodified geometry. The figures on top show Cauchy slices embedded in a high-dimensional fictitious Minkowski geometry, while the figures below are the same geometries illustrated in the stereographic projection, \ie, the Poincaré disk of them, respectively. The figures from left to right represent the AdS Cauchy slices of the original spacetime, the spacetime with IR geometry replaced by a hyperbolic space {with curvature $R_{IR}<R_{AdS}$}, and the hyperbolic extremal case with a negative infinite curvature IR region {$R_{IR}\to -\infty$}, respectively. The boundary subregions are marked by blue and green curves, and their RT surfaces are marked by red curves. In the right figure, the RT surface of the green subregion is undergoing a phase transition to the purple curve.}
    \label{IR-} 
\end{figure}

Similar to the previous case, the RT surface will undergo a phase transition as one enlarges a single boundary subregion. We name the length of this boundary subregion when the phase transition happens the critical length $L_c$. Here note that in the example provided above, we have considered the case where the modified IR geometry has a rotational symmetry. However, in the most general case, we could consider an arbitrarily shaped modified IR region without any symmetry, \ie{} at the edge of the IR region, $r_{IR}$ would depend on the spatial coordinate $x$. In this general case, the critical length $L_c$ would become a function of the position $x$ of the boundary subregion, which is determined by the shape of the IR region.

We now illustrate a geometrical method to determine the critical length $L_c(x)$. As shown in figure \ref{Lc}, the modified IR region is denoted by the upper region bounded by the red curve.  We could have two green dashed horospheres tangential with each other and both tangential to the edge of the IR region. A horosphere is a set of points that are equidistant (at diverging distances) from a point $O$ on the conformal boundary in hyperbolic space. In Poincare coordinates, a horosphere is a round shape that is tangential to the conformal boundary at point $O$. In figure \ref{Lc}, we have the length of $\Gamma_{AM}$ equal to $\Gamma_{AC}$, and the length of $\Gamma_{MB}$ equal to $\Gamma_{BD}$ from the definition of a horosphere. 
Note that, as we have explained, in the hyperbolic extremal IR region, the length of geodesic $\Gamma_{CD}$ tends to zero. Thus, we only need to consider the length of the purple curves after the phase transition.
Therefore, with the length of the blue curve $\Gamma_{AM}+\Gamma_{MB}$ equal to the sum of the lengths of the purple curves $\Gamma_{AC}+\Gamma_{BD}$, the distance between their intersections on the boundary $|AB|$ is the critical distance at this location.
\begin{figure}[H]
    \centering \includegraphics[width=0.6\textwidth]{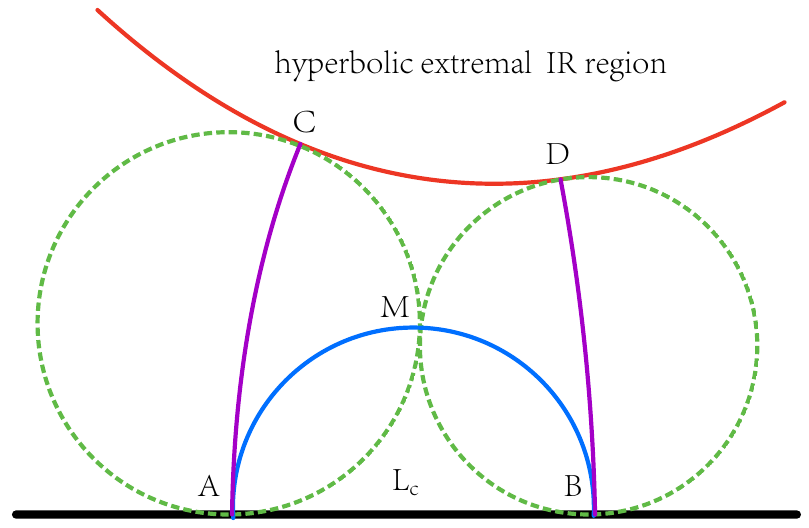} 
    \caption{The phase transition of RT surfaces (from the blue curve to the purple curves) at the critical length $L_c$ {in Poincaré coordinate}. The modified IR region of this hyperbolic extremal case is located above the red curve. The green dashed circles are horospheres which are tangential with each other, and tangential to the boundary of the IR region. According to the definition of the horosphere, the length of the blue curve $\Gamma_{AM}+\Gamma_{MB}$ is equal to the sum of the lengths of the purple curves $\Gamma_{AC}+\Gamma_{BD}$. This results in the line segment $|AB|=L_c$ at the boundary being the critical length.}
    \label{Lc} 
\end{figure}

The RT surface of a boundary subregion with a length greater than the critical length will extend through the IR region. Without loss of generality, when the critical length $L_c(x)$ is a constant function, \ie{} the boundary of the IR region is given by $r=$constant, the RT surface after the phase transition will be two straight lines $\theta=\theta_0$ and $\theta=\theta_0+\frac{L}{l_{S^1}}$. The holographic entanglement entropy will not increase further as one enlarges the length of the boundary subregion $L>L_c$ after the phase transition as the lengths of the purple curves are all equal when $L_c$ is constant. The fact that the entanglement entropy reaches its saturation value at $L=L_c$ indicates that no long-range entanglement between boundary degrees of freedom at a distance $L>L_c$ exists while only short-range entanglement between degrees of freedom  $L<L_c$ near the edge of the boundary subregion contributes to the entanglement. We will provide proof to this conclusion based on the conditional mutual information between subregions in the next section\footnote{
{If $L_c(x)$ varies with $x$,  the same physical interpretation can be made in this more general case.}}.

The radius\footnote{The center of a horosphere is located on the conformal boundary. The radius of a horosphere is defined as the distance between a point on the horosphere and its center, represented by the length of the purple curves in figure \ref{Lc}.} of the horosphere tangential to the edge of the IR region holds specific physical significance. 
It represents the entanglement wedge cross section, half of the mutual information and the squashed entanglement (as described in \cite{Umemoto_2018}) between boundary subregions $A$ and $B$ on their respective left and right sides, with each subregion's length exceeding the critical length $L_c(x)$. 

{The entanglement entropy for a single region with size L on this hyperbolic extremal modified geometry (\ref{Hy}) could be summarized as follows}
\begin{equation}\label{HYP}
    S(L)=
        \left\{
        \begin{aligned}
        &\frac c3 \log(\frac{l_{S^1}}{\pi\epsilon}\sin(\frac{L}{2l_{S^1}}))\quad &(L<L_c\,\text{or}\,L>2\pi l_{S^1}-L_c)\\
        &\frac c3 \log(\frac{l_{S^1}}{\pi\epsilon}\sin(\frac{L_c}{2l_{S^1}})) &(L_c\leq L\leq 2\pi l_{S^1}-L_c).
        \end{aligned}
        \right.
\end{equation}
where the relationship between the critical length $L_c$ and $r_{IR}$ is
\begin{equation}
    \log(\frac{l_{S^1}}{\pi\epsilon}\sin(\frac{L_c}{2l_{S^1}}))=\log(l_{S^1}+\sqrt{1+l_{S^1}^2})-\log(r_{IR}+\sqrt{1+r_{IR}^2}).
\end{equation}

\subsection{Shape constraints on IR regions in both extremal cases}
\noindent
The shape of the IR region in both cases could be quite general when there is not symmetry constraint. Technically, no geometrical shape constraints should be introduced for the IR region. We can arbitrarily choose a region (or even multiple regions) and use hemispheres or light cones (in the embedding language)  with the same circumferences at the edge to replace them, and then glue their edges to the UV geometry. Note that we do not discuss energy conditions of the matter fields responsible for the IR geometry here. However, if we choose a too ``wiggly" region, RT surfaces can only wrap around the convex hull of them rather than touch its edge of the ``wiggly" part, \ie, the ``wiggly" part of the IR region is ``useless" when calculating the RT formula. In other words, one can always ``complete" a very wiggle IR to a larger region while preserving the shape of all the RT surfaces. As shown in figure \ref{wrap}, in the spherical extremal case, the RT formula for a concave IR region is equivalent to the RT formula of the spherical extremal case where the IR region is its convex hull. In the hyperbolic extremal case, the RT formula for a horospherically concave IR region is equivalent to the RT formula of the hyperbolic extremal case where the IR region is its horospherically convex hull. As a result, we demand the IR region to be convex in the spherical extremal case and horospherically convex in the hyperbolic extremal case for this reason. We leave the investigation on the physical consequences of ``wiggly" IR regions to future work.  
\begin{figure}[H]
    \centering \includegraphics[width=1\textwidth]{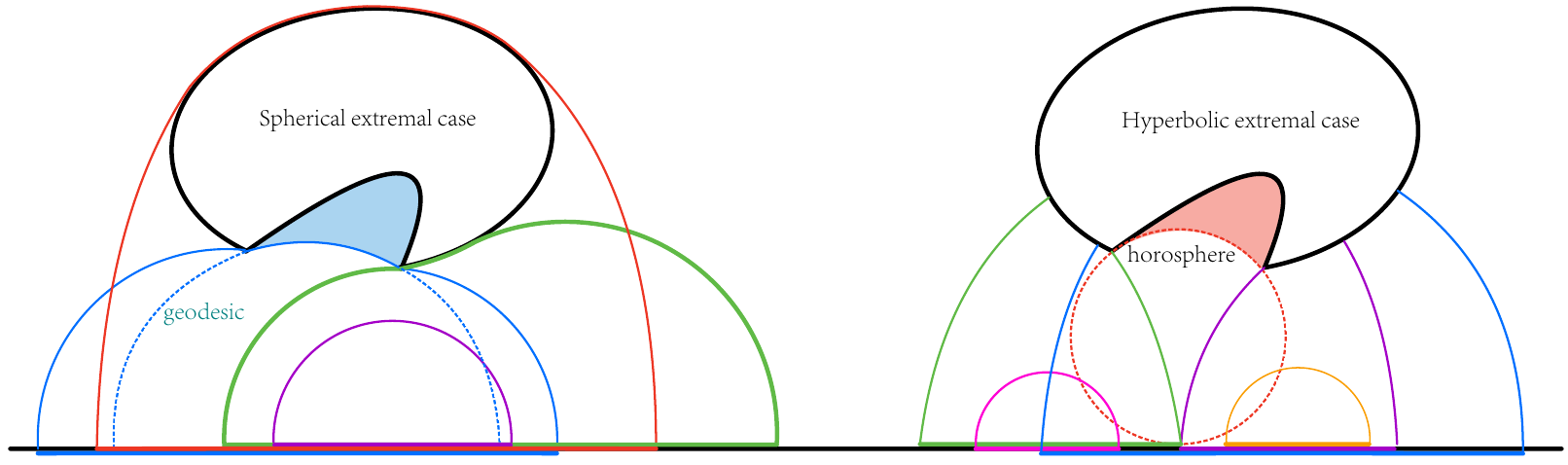} 
    \caption{Shape constraints for the IR regions (denoted by the regions bounded by black curves) in extremal cases. RT surfaces are indicated by colorful curves {in Poincaré coordinate}. Left: in the spherical extremal case, RT surfaces cannot penetrate into the blue region, \ie, the so-called ``wiggly" part of the IR region. Consequently, if we choose its convex hull as the IR region instead, in the spherical extremal case, there would not be any difference between these two cases in terms of holographic entanglement entropy. Right: similarly, in the hyperbolic extremal case, if the IR region is too wiggly, selecting its ``horospherically convex hull" as the IR region instead would not alter the RT surfaces.}
    \label{wrap} 
\end{figure}

\section{Entanglement structure}
\noindent
In this section, based on the formulas of single region entanglement entropy given in the previous section, we analyze the entanglement structure {using RT formula, mutual information, conditional mutual information and the PEE formula}  of the two extreme cases separately.
\subsection{Mutual information}
\noindent
{As proposed in section 1, we anticipate that the entanglement structure, particularly the long-range entanglement, will be modified due to the modification of the IR geometry in the bulk. Mutual information between distant boundary subregions serves as a first measure, {though it is not a very accurate measure as it includes the contribution from both classical correlations and quantum entanglement}. The nonzero mutual information between disconnected boundary subregions is reflected in the connectivity of the entanglement wedge of their union. Modifying the IR geometry into the spherical extremal (or hyperbolic extremal) case will alter the length of geodesics near the IR region, making it more difficult (or easier) to connect the entanglement wedges of disconnected regions.}
\begin{figure}[H]
    \centering \includegraphics[width=0.95\textwidth]{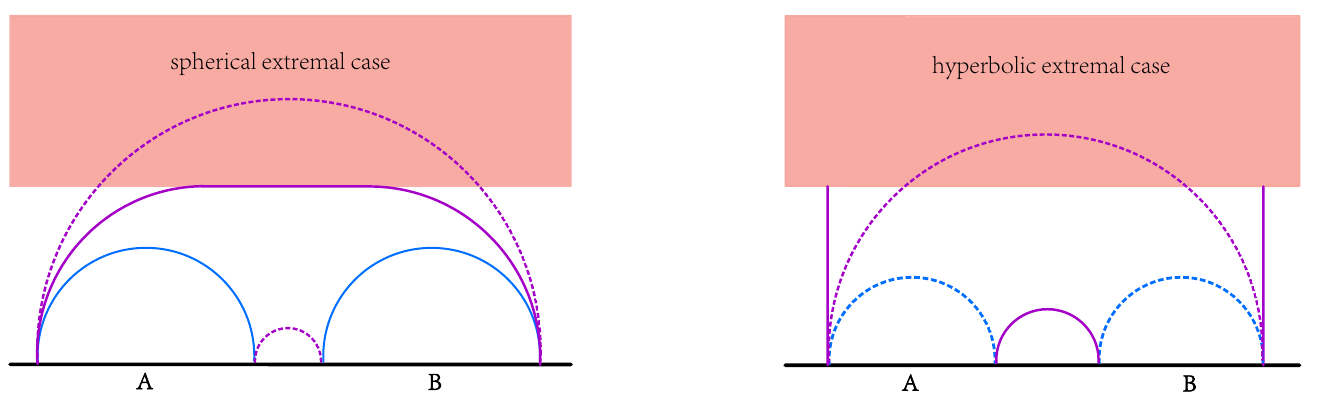} 
    \caption{The entanglement wedges of the union of disconnected boundary intervals $A$ and $B$ are shown for the spherical extremal case (left) and the hyperbolic extremal case (right). The IR region is represented by the red shaded area. For the spherical extremal case, $AB$ is chosen so that its entanglement wedge is connected (bounded by purple dashed curves) before modifying the IR geometry, and disconnected after the modification (bounded by blue curves). This change occurs because the area of the connected RT surface is enlarged by the modified geometry (purple curve). For the hyperbolic extremal case, $AB$ is chosen so that its entanglement wedge is disconnected (bounded by blue dashed curves) before modifying the IR geometry, and connected after the modification (bounded by purple curves).}
    \label{MI} 
\end{figure}

In the spherical extremal case, it is easier for the entanglement wedge of two disconnected regions A and B to be disconnected compared to the unmodified geometry, while in the hyperbolic extremal case, it is the opposite. The phase transition from disconnected entanglement wedge to connected ones becomes more difficult when we decrease the distance between two intervals with fixed length in the spherical extremal case and opposite for the hyperbolic extremal case.  Figure \ref{MI} demonstrates examples where the mutual information is decreased (or enlarged) after modifying the geometry of the IR region (red shaded) into the spherical (hyperbolic) extremal case. This suggests that the {correlation} between $A$ and $B$ in the figure is decreased in the spherical extremal case and enhanced in the hyperbolic extremal case. Moreover, when the distance between $A$ and $B$ is larger than $L_c$, and each has a finite length in Poincaré half-plane coordinates (or $L_A+L_B<\pi l_{S^1}$ in global coordinates), the entanglement wedge of their union must be disconnected in both cases. This suggests the vanishing of {correlation (at $O(N^2)$ order)} for distances longer than $L_c$ in both cases. In the following sections, we can find this general intuition consistent with the fine structure analysis.

However, analyzing only the mutual information is not sufficient for the following reasons. Firstly, mutual information heavily relies on the specific properties of the subregions, e.g. {mutual information always vanishes\footnote{{Note that we are only considering the leading order $O(N^2)$ contributions using the RT formula instead of the Quantum Extremal Surface \cite{Engelhardt:2014gca} formula. Therefore, in the case when $O(N^2)$ order of the mutual information vanishes, the $O(1)$ order correlation does not need to vanish.}} }when the length of one of the boundary intervals is sufficiently small. This limitation prevents it from serving as a precise measure for {correlation} at each length scale. Secondly, mutual information is uniquely determined by $\rho_{AB}$ without revealing how $\rho_{AB}$ is ``embedded" into the entire boundary state. In this sense, it only measures the bipartite {correlation} between $AB$, while overlooking the presence of global (multi-partite) {correlation} between $AB$ and another system $E$. In comparison, conditional mutual information in the next section serves as a more fine entanglement measure.

\subsection{{Conditional mutual information}}
\noindent
In this subsection, we analyze the saturation of strong subadditivity and weak monotonicity for boundary states, \ie, the vanishing condition for conditional mutual information, for the spherical and hyperbolic extremal cases. As will be shown, these saturations imply the elimination of entanglement at specific length scales, paving the way for the analysis of fine entanglement structures.
\subsubsection{The spherical extremal case}
\noindent
Strong subadditivity \cite{Lieb:1973cp}, a fundamental theorem in quantum information theory, states that for quantum systems $A$, $E$, and $B$, we have
\begin{equation}
    S_{AE}+S_{BE}\geq S_{ABE}+S_{E}.
\end{equation}
It has been proven \cite{Hayden_2004} that a necessary (but not sufficient) condition for this equality to be saturated is that $\rho_{AB}$ is a separable state
\begin{equation}
    \rho_{A B}=\sum_j p_j \rho_{A_j} \otimes \rho_{B_j},
\end{equation}
with no quantum entanglement between $A$ and $B$. The conditional mutual information $I(A:B|E)$ is defined by 
\begin{equation}\label{cmi}
    I(A:B|E)=S_{AE}+S_{BE}-S_{ABE}-S_{E},
\end{equation} and its non-negativeness is guaranteed by the strong subadditivity.

For the IR modified geometries in the previous section, in the non-extremal cases, the RT surfaces of $AE$, $BE$, $ABE$ and $E$ do not coincide with each other, as illustrated in figure \ref{CMI}, resulting in a non-zero value of the conditional mutual information from (\ref{cmi})\footnote{Note that we have chosen $E$ to be the region between $A$ and $B$ as shown in figure \ref{CMI}. The choice of $E$ is crucial for the physics discussed here. Different choices of $E$ will result in different values of $I(A:B|E)$, e.g. if $E$ has no quantum entanglement with $AB$, the conditional mutual information will reduce to the mutual information and could not provide extra information about the quantum entanglement structure between $A$ and $B$. Therefore, $E$ should be chosen to have substantial connection with both $A$ and $B$, and the most meaningful and physical choice of $E$ should be the connected region adjacent to $A$ and $B$, otherwise, it will not contain all the global entanglement. The physical significance of this choice of $E$ in conditional mutual information is also evidenced in studies on differential entropy and the kinematic space \cite{Balasubramanian:2013lsa,czech2015integral,Czech_2016}.}. However, this situation changes {in extremal cases}. In the spherical extremal case described above, the RT surfaces of $AE$, $BE$, $ABE$ and $E$ coincide (figure \ref{CMI}) {when and only when the following two conditions are both satisfied: the length of the boundary subregion $E$ is longer than $L_c$, and the length of the boundary subregion $AEB$ is not longer than $\pi l_{S^1}$}.
Therefore, the conditional mutual information vanishes, \ie, $I(A:B|E) = 0$ under these conditions. In other words, $I(A:B|E)=0$ iff the distance $l_{ab}$ between any two points $a\in A$ and $b\in B$ satisfies $L_c\leq l_{ab}<\pi l_{S^1}$.
This result strongly signifies the absence of long-range entanglement ($L_c < L < \pi l_{S^1}$) between boundary subregions $A$ and $B$ and the presence of the longest-range entanglement at $L = \pi l_{S^1}$.

\subsubsection*{Relationship with differential entropy and the generalized Rindler wedge}
\noindent
In \cite{Balasubramanian:2013rqa}, it is shown that the length of a convex curve in the bulk corresponds to the differential entropy on the boundary, which is a linear combination of the entropy of boundary subregions. From the discussion in section 2, one can find that in the spherical extremal case, the part of the RT surface, which wraps around (left of figure \ref{wrap} and right of figure \ref{CMI}) the IR region, is a convex curve \cite{Ju:2023bjl}. {As a result, the holographic entanglement entropy in the spherical extremal case should be described as a form of differential entropy \cite{Balasubramanian:2013rqa} for boundary subregions with length $L_c<L<\pi l_{S^1}-L_c$, due to the fact that the RT surface coincides with the bulk curve responsible for the differential entropy in this case.} In \cite{Czech:2014tva}, the authors define a state whose long-range correlations are purely Markovian\footnote{{A tripartite state $\tilde{\rho}_{AEB}$ is Markovian if its reduced density matrix satisfies $\log \tilde{\rho}_{AEB} = \log \tilde{\rho}_{AE} + \log \tilde{\rho}_{EB} - \log \tilde{\rho}_{E}$, which is equivalent to the condition of zero conditional mutual information, i.e., $I(A:B|E) = 0$. Given this fact, it was shown that the entanglement entropy of the Markovian state should be the differential entropy due to the vanishing conditional mutual information  \cite{Czech:2014tva} as long as the state is a consistent one.
}} with its entanglement entropy given by differential entropy formula. This is consistent with the spherical extremal case we have above, where we have just shown that the long-range correlations do not exist.

These are both closely related to the ``generalized Rindler wedge" in \cite{Ju:2023bjl}, which is the general spacetime subregion accessible to a family of accelerating observers. For physical consistency, the enclosing surface of a generalized Rindler wedge is required to be convex.
In the spherical extremal case, for boundary subregions with RT surfaces in phases 1 and 2 (in the right of figure \ref{CMI}) as described above, their entanglement wedges coincide with the generalized Rindler wedge, thus making the state on them the holographic correspondence of a generalized Rindler wedge in \cite{Ju:2023bjl}. However, since the entire boundary is a pure state here, for boundary subregions in phase 3 or 4, or for the entire boundary, the state on them does not correspond to a generalized Rindler wedge in the bulk.

\subsubsection{The hyperbolic extremal case}
\noindent
\subsubsection*{Strong subadditivity and weak monotonicity}
\noindent
Differential entropy \cite{Balasubramanian:2013lsa,Ju:2023bjl,Ju:2023dzo} (equivalent to the entanglement entropy of the spherical extremal case) is considered the theoretical upper bound for entanglement entropy in boundary subregions, assuming the UV entanglement structure is fixed, due to strong subadditivity. As the diametrically opposed extremal case from a geometrical standpoint, we expect the hyperbolic extremal case to manifest opposite characteristics in its own right. This prompts us to examine another crucial entropy inequality additionally: weak monotonicity, the purified version of strong subadditivity. In the following, we will analyze the strong subadditivity and weak monotonicity in the hyperbolic extremal case.

As shown in figure \ref{SSA}, when the length of region $E$ exceeds or equals the critical length $L_c$, the conditional mutual information takes the form
\begin{equation}
    I(A:B|E)=S_{AE}+S_{BE}-S_{ABE}-S_{E}=a+c+b+d-a-d-b-c=0.
\end{equation}
This implies that the strong subadditivity is saturated, signifying the absence of quantum entanglement between $A$ and $B$. In summary, the hyperbolic extremal case eliminates all quantum entanglement between boundary subregions with a distance longer than the critical length. Unlike the spherical extremal case, here we only require the distance between $A$ and $B$ to be larger than $L_c$; the length of $ABE$ no longer needs to be smaller than $\pi l_{S^1}$. This leads to the vanishing of entanglement longer than $L_c$, including the longest entanglement $L=\pi l_{S^1}$ between antipodal points, in comparison to the spherical extremal case, where the longest entanglement between antipodal points is enhanced.

\begin{figure}[H]
    \centering \includegraphics[width=0.95\textwidth]{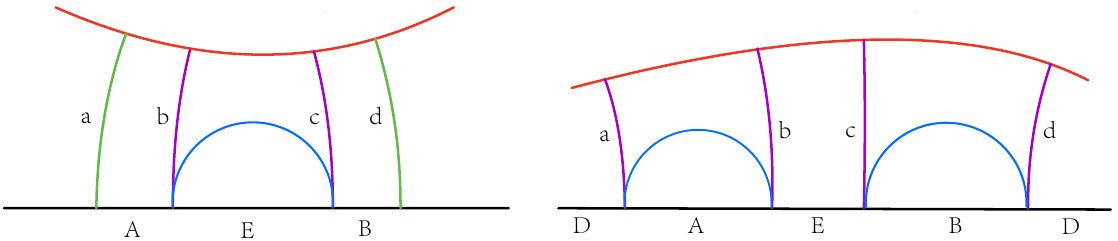} 
    \caption{In the hyperbolic extremal case, we illustrate two scenarios where strong subadditivity and weak monotonicity are saturated separately. Left: strong subadditivity is saturated. The length of region $E$ equals the critical length $L_c$, and its RT surface is marked by purple curves $b$ and $c$. The RT surface for region $ABE$ is marked by green curves. Right: weak monotonicity is saturated. The lengths of both regions $A$ and $B$ reach the critical length $L_c$, with their RT surfaces indicated by purple curves.}
    \label{SSA} 
\end{figure}
Weak monotonicity can be derived from the strong subadditivity, with the introduction of an extra system $D$ such that $ABDE$ is a pure state, \ie
\begin{equation}
    \begin{aligned}
        S_{AE}+S_{AD}&\geq S_{A}+S_{ADE},\\
        \Rightarrow S_{AE}+S_{BE}&\geq S_{A}+S_{B}.
    \end{aligned}
\end{equation}
It could be found that the necessary condition for these inequalities to be saturated is that $\rho_{DE}$ is a separable state on $D$ and $E$.

For the non-extremal modified geometries, the weak monotonicity cannot be saturated due to the geometrical reasons the same as the one for strong subadditivity \cite{Headrick:2007km}. However, in the hyperbolic extremal case, this situation changes. As illustrated in figure \ref{SSA}, when the lengths of regions $A$ and $B$ reach the critical length $L_c$, we have
\begin{equation}
    S_{AE}+S_{BE}=a+c+b+d=S_A+S_B.
\end{equation}
This saturation aligns with the elimination of quantum entanglement between $D$ and $E$, as the separation distance between $D$ and $E$ exceeds the critical length on both sides of $E$.

In contrast to the spherical extremal case, which causes the entanglement entropy of the boundary subregion as $S_{ABE}$ to reach the upper bound due to the saturation of the strong subadditivity, the hyperbolic extremal case makes the sum of the entanglement entropy of two boundary subregions, $AE$ and $BE$, reach the lower bound $S_A+S_B$ due to the saturation of the weak monotonicity. Similar to the spherical extremal case, where the saturation of strong subadditivity implies the elimination of long-range entanglement between $A$ and $B$, here the saturation of weak monotonicity in the hyperbolic extremal case indicates the elimination of entanglement between $D$ and $E$. 

{In a recent paper \cite{Harper:2024aku}, the authors present a proof of the SSA and monogamy of mutual information (MMI) in scenarios where EoW branes are present in the bulk. Within our framework, the presence of EoW branes is equivalent to an extremal form of modified IR geometry in the sense of entanglement entropy. This interpretation facilitates the straightforward application of the MMI proof \cite{Hayden:2011ag} to general spacetime geometries, including those with EoW branes (or spherical extremal IR regions), aligning with the findings in \cite{Harper:2024aku}. Furthermore, it is not only the MMI but also all other entropy inequalities examined in the studies of the holographic entropy cone \cite{Hernandez-Cuenca:2023iqh,Bao:2015bfa,HernandezCuenca:2019wgh} that are fulfilled for \(AdS_{d+1}/BCFT_d\) cases in a similar manner.

In the next section, we will analyze deeper into the quantitative modifications in the entanglement structure utilizing the tools of partial entanglement entropy, confirming the consistency of these qualitative results obtained from the conditional mutual information.

\subsection{Fine entanglement structures}
\noindent
From the analysis of the conditional mutual information, we can intuitively see that in the spherical extremal case, entanglement with a range $L_c < L < \pi l_{S^1}$ is eliminated, and the entanglement with the longest range $L =\pi l_{S^1}$ is enhanced. In the hyperbolic extremal case, entanglement with a range $L > L_c$ is eliminated, and the phase transition of the RT surface implies some nontrivial entanglement structures at the critical length $L_c$. This motivates us to employ tools beyond the RT formula and conditional mutual information, such as the {partial entanglement entropy (PEE)} formula, to further analyze the fine entanglement structures and verify this intuition, as well as compare the differences before and after the IR geometry is modified. For simplicity, we will only focus on the case with spherical symmetry (\ref{ansatz}), while more general shaped IR regions will be explored in future work \cite{Ju:2024}.
\subsubsection{A brief review of PEE}
\noindent
PEE and entanglement contour \cite{Freedman:2016zud,Wen:2018whg,Wen:2019iyq,Wen:2020ech,Kudler-Flam:2019oru,Lin:2022aqf,Lin:2021hqs,Vidal:2014aal,Lin:2023rbd}, denoted as $s_A(A_i)$ and $f_{\mathcal{A}}(\mathbf{x})$, provides an answer to the question of ``how much does $A_i\subset A$ contribute to the entanglement entropy of $A$?" By definition, we have
\begin{equation}
\begin{aligned}
    s_{\mathcal{A}}(A_i)=\int_{\mathcal{A}_i} f_{\mathcal{A}}(\mathbf{x}) d \sigma_{\mathbf{x}},
\end{aligned}
\end{equation}
where $\mathbf{x}$ denotes a point in $\mathcal{A}$, $\sigma_{\mathbf{x}}$ denotes an infinitesimal subset of $\mathcal{A}$ at $\mathbf{x}$, and the normalization condition is
\begin{equation}
    S(A)=\int_{\mathcal{A_i}} f_{\mathcal{A}}(\mathbf{x}) d \sigma_{\mathbf{x}}.
\end{equation}
{The physical consistency of PEE requires uniqueness of its definition \cite{Kudler-Flam:2019nhr}.}
Although we can establish reasonable constraints on PEE function \cite{Vidal:2014aal}, it cannot be uniquely determined. As a solution, one can introduce a permutation condition as an additional constraint \cite{Wen:2019iyq}, or directly utilize the so-called PEE$=$CMI {(Conditional Mutual Information)} proposal or  as demonstrated, which leads to the same consistent result as follows
\begin{equation}\label{PEEF}
s_A(A_i)=\frac{1}{2}[  S(A_i \mid A_1 \cup \cdots \cup A_{i-1}) +S(A_i \mid A_{i+1} \cup \cdots \cup A_n)],
\end{equation}
where $A_1, \ldots, A_n$ are put in order from left to right composing a connected region A. It is exactly half of the conditional mutual information $I(A_i:\bar{A}|A_1 \cup \cdots \cup A_{i-1})$, {which could be summarized as ``PEE$=$CMI".}

An equivalent way to rewrite this formula \cite{Wen:2019iyq,Lin:2023rbd} is to use the two-point PEE function $\mathcal{I}(\mathbf{x}, \mathbf{y})$ to describe the ``density of entanglement" between two points. {More explicitly, $\mathcal{I}(\mathbf{x}, \mathbf{y})$ quantifies the partial entanglement entropy between $A$ and $B$ contributed by $d\sigma_x\subset A$ and $d\sigma_y\subset B$.} Thus we have \footnote{{
Note that the definitions of the partial entanglement entropy from equations (\ref{Normalize}) and (\ref{highdimpee}) look similar to the expression for the extensive mutual information model in \cite{Casini:2008wt} 
\begin{equation*} I(A,B)=\int_A d \sigma_x \int_B d\sigma_y j(x,y,\eta_{A,x},\eta_{B,y}).
\end{equation*} However, these two are different in nature. In (\ref{Normalize}), the left side is $s_{\bar{B}}(A)$, which is different from $I(A,B)$ because $s_{\bar{B}}(A)$ refers to the amount of entanglement that $A$ contributes to the total amount of entanglement of $B$ (or equivalently $\bar{B}$). $s_{\bar{B}}(A)$ is in general not $ I(A,B)$ (while it is $I(A,B|E)$ with E being the region between $A$ and $B$ as we will explain below). Equation (\ref{Normalize}) could be thought as the definition of $s_{\bar{B}}(A)$  with the right-hand side $j$-function determined from several constraints. In 2+1 dimensions,  the $j$-function in equations (\ref{Normalize}) and (\ref{highdimpee}) is the conditional mutual information of two infinitesimal subregions of $\sigma_x$ in $A$ and $\sigma_y$ in $B$. The conditional mutual information has by definition a chain rule 
\begin{equation*} 
    I(A,CD|B)=I(A,C|B)+I(A,D|BC), 
\end{equation*} which ensures that the integration on the whole regions of $A$ and $B=\bar{A}$ in equation (\ref{Normalize}) would give the entanglement entropy of $S(A)$. This chain rule of the conditional mutual information holds by definition, and does not require the mutual information to be extensive. Therefore, though looking similar, the equations (\ref{Normalize}) and (\ref{highdimpee}) are different in nature from that in \cite{Casini:2008wt} and the $j$ function in \cite{Casini:2008wt} is in general not the conditional mutual information as in our case. These two are only equal when the mutual information is extensive, i.e. when $ I(A,C|B)=I(A,C)$ \cite{Agon:2021zvp}, which, however, we do not require and is not true in the holographic case.}
}
\begin{equation}\label{Normalize}
s_{\bar{B}}({A})=\int_A \mathrm{~d} \sigma_{\mathbf{x}} \int_B \mathrm{~d} \sigma_{\mathbf{y}} \mathcal{I}(\mathbf{x}, \mathbf{y}),
\end{equation}
where $s_{A}(A)=S(A)$ is the normalization condition \cite{Wen:2019iyq} of PEE.

{In a CFT, the two-point PEE function $\mathcal{I}(\mathbf{x}, \mathbf{y})$ could be greatly constrained by conformal symmetry, up to a single undetermined constant as follows\footnote{For ball-shaped regions in general dimensions, this formula coincides with the PEE=CMI formula. However, the physical validity of this procedure for regions of arbitrary shapes in higher-dimensional CFTs, where 
$d>2$, is questionable. Nevertheless, since it aligns with the PEE formula in the case of 
$d=2$, which is the focus of our subsequent analysis, its validity in this specific context is assured.}} 
\begin{equation}\label{highdimpee}
\mathcal{I}\left(\mathbf{x}, \mathbf{y}\right)\propto \frac1{|\mathbf{x}-\mathbf{y}|^{2(d-1)}}.
\end{equation}
This formula is invalid for cases that we are analyzing, as the conformal symmetry is broken when the IR geometry is modified in the bulk. However, we could determine $\mathcal{I}\left(\mathbf{x}, \mathbf{y}\right)$ using the PEE$=$CMI proposal or equivalently the formula (\ref{PEEF}). Here, without loss of generality, we consider the simplest case where the IR region is spherically symmetrical in the Poincaré disk (figure \ref{IR-}), so that the two-point PEE function is still translational invariant (or rotational invariant in finite size CFT), which is solely a function of the distance $L=|\mathbf{x}-\mathbf{y}|$. 
Utilizing the PEE formula (\ref{PEEF}), we have
\begin{equation}
\begin{aligned}
    \mathcal{I}(L)d\sigma_xd\sigma_y&=\frac12I(d\sigma _x:d\sigma_y|E)
    \\&=\frac{1}{2}(S_{d\sigma_x E}+S_{d\sigma_y E}-S_{d\sigma_x Ed\sigma_x }-S_E)\\&=\frac12 (S(L+d\sigma_x)+S(L+d\sigma_y)-S(L+d\sigma_x+d\sigma_y)-S(L))\\&=-\frac{1}{2}\frac{d^2S(L)}{dL^2}d\sigma_xd\sigma_y,
\end{aligned}
\end{equation}
where $S(L)$ represents the entanglement entropy of a boundary subregion with a length of $L$, \ie{}, region $E$ between $d\sigma_x$ and $d\sigma_y$. The positivity of $\mathcal{I}(L)$ is a result of the convexity of the von Neumann entropy.
In the language of integral geometry \cite{czech2015integral,Czech_2016}, $\mathcal{I}(L)$ is the density of geodesics (Crofton density) which could be interpreted as conditional mutual information physically. This interpretation is consistent with the PEE proposal mentioned above, and the normalization condition (\ref{Normalize}) could be verified as follows
\begin{equation}
    s_A(A)=\int_0^{L_A} dx(\int^0_{-\infty} \mathcal{I}(|x-y|)dy +\int_{L_A}^\infty\mathcal{I}(|x-y|)dy)=S(L_A),
\end{equation}
where $L_A$ represents the length of $A$, and $\bar{A}$ spans an infinitely large region of $(-\infty,0)\cup(L_A,\infty)$.

It is worth noting that the PEE formula becomes invalid for disconnected region $A$ on the boundary \cite{Lin:2023orb}, and a natural explanation is that there are multipartite entanglement in this system, which should be treated carefully for entanglement in disconnected regions. As a consequence, our conclusions and figures below cannot be used for disconnected intervals on the boundary. 
\subsubsection{The spherical extremal case}
\noindent
In the spherical extremal case, the entanglement entropy of a boundary subregion is given by (\ref{SPH}), which we rewrite here for convenience
\begin{equation}
    S_{sph}(L)=
        \left\{
        \begin{aligned}
        &S_{vac}=\frac c3 \log(\frac{l_{S^1}}{\pi\epsilon}\sin(\frac{L}{2l_{S^1}})),\,\,\,\,\quad\quad\quad\quad\quad\quad\quad\quad\quad\quad(L<L_c\,\text{or}\,L>2\pi l_{S^1}-L_c)\\
        &\frac c3 \log(\frac{l_{S^1}}{\pi\epsilon}\sin(\frac{L_c}{2l_{S^1}}))+\frac {c\cot(L_c/2l_{S^1})}{6l_{S^1}}(L-L_c) ,\,\,\quad\quad\quad\quad\quad\quad\quad(L_c\leq L\leq \pi l_{S^1})\\
        &\frac {c}3 \log(\frac{l_{S^1}}{\pi\epsilon}\sin(\frac{L_c}{2l_{S^1}}))-\frac {c\cot(L_c/2l_{S^1})}{6l_{S^1}}(L-2\pi l_{S^1}+L_c), \,( \pi l_{S^1} \leq L\leq 2\pi l_{S^1}-L_c)
        \end{aligned}
        \right.
\end{equation}
{where $S_{vac}$ is the entanglement entropy of the boundary subregion before the IR geometry is modified.} The two-point PEE function could be calculated as
\begin{equation}\label{2PEEsph}
    \mathcal{I}_{sph}(L)=\left\{
        \begin{aligned}
        &\mathcal{I}_{vac}=\frac {c}{24} \frac{\csc(L/2l_{S^1})^2}{l_{S^1}^2},\,\,\,\,\,\,\quad\quad \quad \quad\quad\,\,\,\quad \quad\quad \quad(L<L_c\,\,\text{or}\,\,L>2\pi l_{S^1}-L_c)\\
    &\quad0,\,\,\quad \quad \quad\quad \quad \quad \quad\quad \quad\quad\quad\quad\,\,\, (L_c\leq L<\pi l_{S^1}\,\text{or}\,\pi l_{S^1}< L\leq 2\pi l_{S^1}-L_c)\\
        &\frac {c\cot(L_c/2l_{S^1})}{6l_{S^1}}\delta(L-\pi l_{S^1}) =\delta(L-\pi l_{S^1})\int_{L_c}^{2\pi l_{S^1}-L_c}\mathcal{I}_{vac}(L)dL.\quad (L\rightarrow \pi l_{S^1}),
        \end{aligned}
        \right.
\end{equation}
{where $\mathcal{I}_{vac}$ represents the two-point PEE function before the modification of the IR geometry. 

For a boundary subregion $A$ that spans $[0,L_A]$ with $L_A<L_c$, the PEE $s_{A}(dL)$ for $dL\subset A$ locating at $x=l_0$ could be calculated by integrating $\mathcal{I}_{sph}(L)$ on $\bar{A}$ using formula (\ref{Normalize})
\begin{equation}\label{point}
\begin{aligned}
    s_{A}(dL)&=dL\int_{l_0}^{2\pi l_{S^1}-L_A+l_0}\mathcal{I}_{sph}(l)dl.\\
    &=dL(\int_{l_0}^{L_c}\mathcal{I}_{sph}(l)dl+\int_{L_c}^{2\pi l_{S^1}-L_c}\mathcal{I}_{sph}(l)dl+\int_{2\pi l_{S^1}-L_c}^{2\pi l_{S^1}-L_A+l_0}\mathcal{I}_{sph}(l)dl).
\end{aligned}
\end{equation}
From equation (\ref{2PEEsph}), we observe that the integral $\int_{L_c}^{2\pi l_{S^1}-L_c}\mathcal{I}(L)dL$ remains unchanged upon modifying the IR geometry, also, the first and the third terms of (\ref{point}) remain unchanged trivially as $\mathcal{I}_{sph}(L)=\mathcal{I}_{vac}(L)$ in their regimes. 
Consequently, the PEE $s_{A}(dL)$ remains unchanged, \ie, the contribution to the entanglement entropy $S_A$ from any small subregion $dL$ inside $A$ remains unchanged. As a consequence, the entanglement entropy of a boundary subregion with length $L<L_c$ remains the same after modifying the IR geometry.

In contrast, the entanglement entropy of boundary subregions with $L_c<L<2\pi l_{S^1}-L_c$ would change. Consequently, the unchanged $\int_{L_c}^{2\pi l_{S^1}-L_c}\mathcal{I}(L)dL$ indicates that the modification of IR geometry results solely in a redistribution of entanglement with range $L\in [L_c,2\pi l_{S^1}-L_c]$.}
We could observe that the two-point PEE diverges at $L=\pi l_{S^1}$ where the phase transition occurs from phase II to phase III, while it vanishes at other values of $L\in [L_c,2\pi l_{S^1}-L_c]$.
This result can be understood as a clear indication that the entanglement structure with distances longer than the critical distance $L>L_c$ is eliminated and instead ``transfers" to the longest length entanglement at $L=\pi l_{S^1}$ (antipodal points on the boundary $S^1$). This explicit calculation of fine entanglement structure coincides with the intuition we obtain from analyzing RT formula and CMI in the previous sections.

\subsubsection*{Entanglement capacity of the IR region}
\noindent
From the discussion above, we can conclude that modifying the IR geometry alters exclusively the long-range entanglement structure. This leads to an intriguing question: how much long-range entanglement can be changed at most, given a definite IR region to be modified? This maximum represents the carrying capacity of a bulk spatial subregion for the entanglement on the boundary. It is expected to be finite, as long-range entanglement does not possess UV divergence. This value can be readily determined in the spherical extremal case, where all long-range entanglement (referring to entanglement at real space scales $L\geq L_c$) that could be modified is transferred to the longest range. 

Consequently, one can calculate this value by integrating these divergent `PEE pairs' at longest range in the spherical extremal case as follows 
\begin{equation}\label{areaterm}
    \pi l_{S^1}\int_{\pi l_{S^1}-\epsilon}^{\pi l_{S^1}+\epsilon}\mathcal{I}_{sph}(L)dL=\pi \frac c6\cot(L_c/2l_{S^1}),
\end{equation}
where $\epsilon$ is a small positive constant, and the coefficient $\pi l_{S^1}$ only counts half of the CFT circle $S^1$, as we are considering the longest entanglement between this half and its complement. Note that (\ref{areaterm}) is the integral of the divergent part of the two point PEE function, which, different from the total entanglement entropy of half space, isolates the longest range entanglement for the half space. Therefore, (\ref{areaterm}) gives the entanglement entropy of the longest range entanglement for half space. Compared with the $L=\pi l_{S^1}$ result in (\ref{SPH}) of the total entanglement entropy of half space, we could see that this corresponds to the $\frac{c\cot(L_c/2l_{S^1})}{6l_{S^1}}L$ term, while the other two terms are exactly the shorter range entanglement.

Recall that the radius coordinate $ r_{\text{IR}}$ corresponds to the deepest point \( P \) on the RT surface associated with the boundary subregion of length \( L_c \) (see equation (\ref{deepest})), so we have \( \cot\left({L_c}/{2l_{S^1}}\right) = r_{\text{IR}} \). Therefore, the value \( 2\pi \cot\left({L_c}/{2l_{S^1}}\right) \) represents the area of the IR region, and (\ref{areaterm}) becomes
\begin{equation}\label{AREA}
    2\int \text{diverge two-point PEE}=\frac 1{4G}\text{Area}(\text{IR region}),
\end{equation}
where $c=3l_{AdS}/2G$.
Consequently, in this spherical extremal case, we can conclude that the carrying capacity of total entanglement for the IR region is determined by its surface area. This area law provides a new perspective for us to understand the connection between geometry and entanglement. This conclusion can be generalized to the spherical extremal cases of any convex IR region using a specific geometric canceling trick, and {also to general entanglement shadow IR regions}, which we will discuss in detail in \cite{Ju:2024}.

\subsubsection{The hyperbolic extremal case}
\noindent
In the hyperbolic extremal case, the entanglement entropy of a boundary subregion is given by (\ref{HYP}), which we rewrite here for convenience
\begin{equation}
    S_{hyp}(L)=
        \left\{
        \begin{aligned}
        &\frac c3 \log(\frac{l_{S^1}}{\pi\epsilon}\sin(\frac{L}{2l_{S^1}}))\quad\quad\quad\quad (L<L_c\,\text{or}\,L>2\pi l_{S^1}-L_c)\\
        &S_{vac}(L_c) \quad\quad\quad\quad\quad\quad\quad\quad\,\,\,\,\,(L_c\leq L\leq 2\pi l_{S^1}-L_c).
        \end{aligned}
        \right.
\end{equation}
The two-point PEE could be obtained to be
\begin{equation}
    \mathcal{I}_{hyp}(L)=\left\{
        \begin{aligned}
        &\mathcal{I}_{vac}=\frac {c}{24} \frac{\csc(L/l_{S^1})^2}{l_{S^1}^2}\quad\quad \quad\quad\quad\quad\quad\quad\quad\,(L<L_c\,\text{or}\,L>2\pi l_{S^1}-L_c)\\
    &\frac {c\cot(L_c/2l_{S^1})}{12l_{S^1}}\delta(L-L_c)=\int_{L_c}^{\pi l_{S^1}}-\frac12 S_{vac}''(L)dL \delta(L-L_c),\quad (L\rightarrow L_c)\\
        &\quad0\quad \quad \quad \quad \quad \quad\quad\quad\quad\quad\quad\quad\quad\quad\quad\quad \quad \quad \quad (L_c\leq L\leq 2\pi l_{S^1}-L_c),
        \end{aligned}
        \right.
\end{equation} which diverges at the phase transition $L=L_c$.

{Again, the integral $\int_{L_c-\epsilon}^{2\pi l_{S^1}-L_c+\epsilon}\mathcal{I}(L)dL$ remains unchanged upon modifying the IR geometry.} Similar to the spherical extremal case, this implies that the total amount of entanglement at scales of $L\in [L_c,2\pi l_{S^1}-L_c]$ remains unchanged. Consequently, this again indicates that the modification of IR geometry results in a redistribution of entanglement within range $L\in [L_c,2\pi l_{S^1}-L_c]$. The difference is that here the two point PEE diverges at $L=L_c$, which indicates that the entanglement structure with distances longer than the critical distance $L_c$ is eliminated and transferred to the critical length entanglement at $L=L_c$. Figure \ref{IRES} gives an illustration of the entanglement structure before and after the modification for this hyperbolic extremal case.

\begin{figure}[H]
    \centering 
    \includegraphics[width=0.8\textwidth]{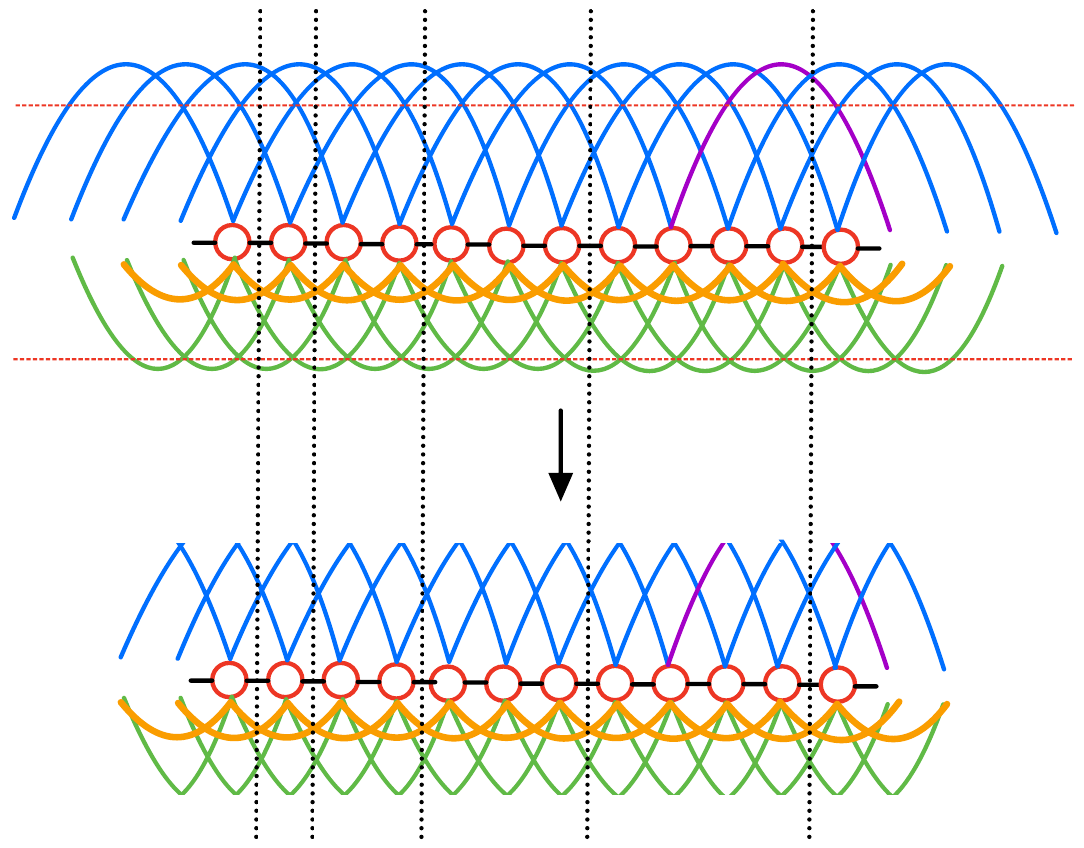} 
    \caption{Entanglement structures before (top) and after (down) the modification of the IR geometry in the hyperbolic extremal case. The boundary system is represented using a combination of equal size discrete subsystems for illustrational convenience, which are indicated by red circles in the figures. Black, orange, green and blue curves represent entanglement between two quantum subsystems at distance scales of 1, 2, 3 and 4, respectively. The black dashed lines exhibit quantum systems with lengths of 1, 2, 3, and 4, whose entanglement entropies are 8, 14, 18, and 20, respectively in the upper figure. The red dashed horizontal lines specify the change of the entanglement structure at $L>L_c$ due to the modification of the IR geometry, \ie, the entanglement structure between the two red dashed lines are not modified while that outside the two lines are modified. The critical length $L_c$ is 2 in this case. After the modification of the IR geometry (lower figure), long-range entanglement with $L>L_c$ is eliminated and transferred to the critical length entanglement as shown by the newly connected blue and green curves. After that, the entanglement entropies of quantum systems with lengths of 1, 2, 3, and 4 are now 8, 14, 14, and 14, respectively.} \label{entstr-}
\end{figure}

\subsubsection*{Connection with EoW brane and AdS/BCFT}
After analyzing the simplest case with translational symmetry, we could easily generalize the shape of the IR regions in this hyperbolic extremal case. The primary change is that the critical length becomes dependent on the location, causing the red dashed line in figure \ref{entstr-} to `curve'. In the special case where the edge of the IR region extends to the boundary and coincides with an EoW brane in the context of AdS/BCFT or brane-world holography, near the intersection of the boundary and the EoW brane, the critical length $L_c$ of the entanglement entropy phase transition decreases to zero (figure \ref{BCFT}). As the entanglement at scales larger than \(L_c\) is completely eliminated, this leads to the direct conclusion that the degrees of freedom extremely close to the boundary of the BCFT can never be entangled with other degrees of freedom at any finite distance.
There should be much more to explore regarding the relation between the hyperbolic case and the AdS/BCFT framework, which we will not discuss here.

\begin{figure}[H]
    \centering \includegraphics[width=0.8\textwidth]{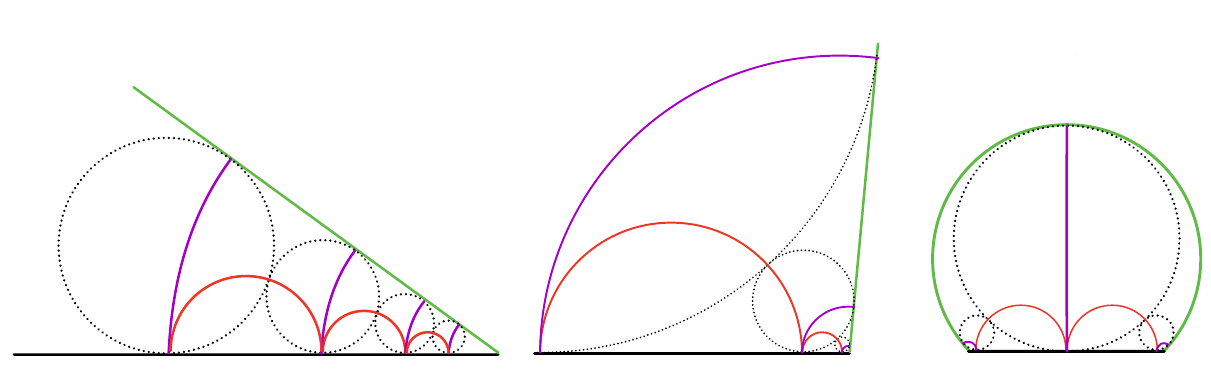} 
    \caption{Critical lengths $L_c$ (denoted by the interval under the red curves) of the boundary system with an EoW brane in the bulk. Those branes have constant tension as in the AdS/BCFT correspondence. EoW brane with negative tension is depicted in the left figure, while those with positive tension are shown in the middle and right figures as green curves in the bulk. Dashed circles represent horospheres, and RT surfaces at the transition point are marked by red and purple curves.} \label{BCFT}
\end{figure}

\section{Conclusion and discussion}
\noindent
In this paper, we propose that modifying IR geometry in the bulk corresponds to modifying the real space long-range entanglement structure of the boundary quantum system. We considered two specific kinds of modified IR geometries: the spherical and hyperbolic IR geometries. In both cases, we analyzed their geometric and RT surface properties. We found that in the spherical extremal case, the IR region becomes an entanglement shadow, which has no RT surface penetrating inside. While in the hyperbolic extremal case, RT surfaces tend to perpendicularly penetrate into the boundary of the IR region. These two extremal cases are closely related to the concepts of differential entropy and brane-world holography, respectively.

We utilized the RT formula, mutual information, conditional mutual information and PEE formula to study the fine entanglement structures of the boundary quantum states. The results indicate a physical picture that modifying the IR geometry is, in fact, a ``redistribution of the entanglement structure at different length scales". We summarize the entanglement structures in figure \ref{IRES}. In the spherical extremal case, long-scale (longer than the critical length) entanglement is transferred to the longest-scale entanglement, while in the hyperbolic extremal case, the long-scale entanglement is transferred to the critical-scale (the shortest scale that could be modified) entanglement at $(L=L_c)$. These diametrically opposed entanglement structures elegantly correspond to the diametrically opposed geometry properties of the spherical extremal case and the hyperbolic extremal case.
\begin{figure}[H]
    \centering \includegraphics[width=0.9\textwidth]{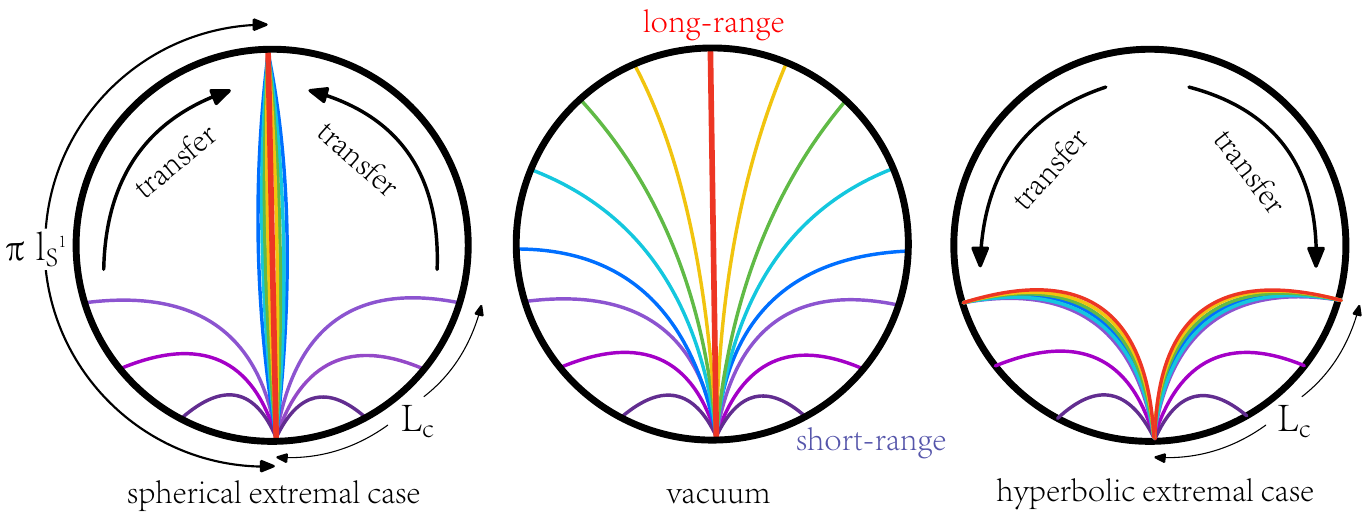} 
    \caption{Entanglement structures are depicted using threads that represent entanglement (two-point PEE function) between the two points that they connect. The middle figure illustrates the entanglement in vacuum AdS, which has entanglement at all length scales. The left figure depicts the spherical extremal case, where all long-scale entanglement at $L > L_c$ is eliminated and is instead transferred to the longest scale $L = \pi l_{S^1}$ entanglement. The right figure shows the hyperbolic extremal case, where all long-scale entanglement with $L > L_c$ is eliminated and is transferred to the critical length $L = L_c$ entanglement.}\label{IRES} 
\end{figure}

An interesting result arises in the spherical extremal case: when we integrate the divergent partial entanglement entropy between the antipodal points, we obtain (\ref{AREA}), where we interpret the area of the boundary of the IR region as the maximum capacity of long-scale entanglement of this IR region. This new interpretation, which will be shown to be valid in more generalized cases in \cite{Ju:2024}, provides fundamental constraints for manipulating entanglement structures when modifying IR geometry. Further physical consequences of modified IR geometries in other quantum informatic properties will be considered in future work, e.g. the entanglement wedge cross-section.

\section*{Acknowledgement}

We would like to thank Yan Liu, Yuan-Tai Wang, Zheng Yan and Yang Zhao for useful discussions.
This work was supported by Project 12347183, 12035016 and 12275275 of the National Natural Science Foundation of China and Grant No. 1222031 of Beijing Natural Science Foundation.  
\appendix

\bibliography{reference}
\bibliographystyle{JHEP}

\end{document}